\documentclass[aps,twocolumn,reprint,pre, superscriptaddress,groupeaddress]{revtex4}

\usepackage{graphpap}
\usepackage[dvips]{graphicx}
\usepackage[dvips]{graphics}
\usepackage{color}
\usepackage{hyperref}
\usepackage{amsmath}
\usepackage[normalem]{ulem}

\graphicspath{{./figures/}}

\begin{document}

\title{Ultrasensitive displacement noise measurement of carbon nanotube mechanical resonators}
\author{S. L. de Bonis}
\altaffiliation{These authors contributed equally to this work.}
\affiliation{ICFO - Institut de Ciencies Fotoniques, The Barcelona Institute of Science and Technology, 08860 Castelldefels, Barcelona,Spain}
\author{C. Urgell}
\altaffiliation{These authors contributed equally to this work.}
\affiliation{ICFO - Institut de Ciencies Fotoniques, The Barcelona Institute of Science and Technology, 08860 Castelldefels, Barcelona,Spain}
\author{W. Yang}
\author{C. Samanta}
\author{A.~Noury}
\altaffiliation{Present address: Laboratoire Charles Coulomb (L2C), Univ Montpellier, CNRS, Montpellier, France.}
\affiliation{ICFO - Institut de Ciencies Fotoniques, The Barcelona Institute of Science and Technology, 08860 Castelldefels, Barcelona,Spain}
\author{J. Vergara-Cruz}
\affiliation{ICFO - Institut de Ciencies Fotoniques, The Barcelona Institute of Science and Technology, 08860 Castelldefels, Barcelona,Spain}
\author{Q. Dong}
\author{Y. Jin}
\affiliation{Centre de Nanosciences et de Nanotechnologies, CNRS, Univ. Paris-Sud, Univ. Paris-Saclay, C2N – Marcoussis, 91460 Marcoussis, France}
\author{A. Bachtold}
\affiliation{ICFO - Institut de Ciencies Fotoniques, The Barcelona Institute of Science and Technology, 08860 Castelldefels, Barcelona,Spain}

\begin{abstract}
Mechanical resonators based on a single carbon nanotube are exceptional sensors of mass and force. The force sensitivity in these ultra-light resonators is often limited by
the noise in the detection of the vibrations. Here, we report on an ultra-sensitive scheme based on a RLC resonator and a low-temperature amplifier to detect nanotube vibrations. We also show a new fabrication process of electromechanical nanotube resonators to reduce the separation between the suspended nanotube and the gate electrode down to $\sim 150$~nm. These advances in detection and fabrication allow us to reach $0.5~\mathrm{pm}/\sqrt{\mathrm{Hz}}$ displacement sensitivity. Thermal vibrations cooled cryogenically at 300~mK are detected with a signal-to-noise ratio as high as 17~dB. We demonstrate $4.3~\mathrm{zN}/\sqrt{\mathrm{Hz}}$ force sensitivity, which is the best force sensitivity achieved thus far with a mechanical resonator. Our work is an important step towards imaging individual nuclear spins and studying the coupling between mechanical vibrations and electrons in different quantum electron transport regimes.

\end{abstract}
\maketitle
The smallest operational mechanical resonators are based on low-dimensional materials, such as carbon nanotubes~\cite{Sazonova2004}, graphene~\cite{Bunch2007,Singh2010,Will2017}, semiconducting nanowires~\cite{Feng2007,Gil-Santos2010,Sansa2012,Montinaro2014}, and levitated particles~\cite{Gieseler2012,Kiesel2013}. Such resonators are fantastic sensors of external forces~\cite{Nichol2008,Moser2013,Lepinay2016,Rossi2016} and the adsorption of mass~\cite{Chiu2008,Wang2010,Chaste2012}. They also provide a versatile platform for fundamental science, including the study of noise~\cite{Gieseler2013a,Zhang2014,Miao2014}, nonlinear phenomena~\cite{Eichler2011a,Deng2016,Alba2016,Mathew2016}, electron-phonon coupling~\cite{Lassagne2009,Steele2009,Ganzhorn2012,Benyamini2014,Ares2016}, and light-matter interaction~\cite{Gloppe2014,Reserbat-Plantey2016}. The greatest challenge with these tiny resonators is to transduce their mechanical vibrations into a measurable electrical or optical output signal. Novel detection methods have been continuously developed over the years~\cite{Purcell2002,Chen2009,Huttel2009,Gouttenoire2010,Arcizet2011,Moser2014,Stapfner2013,Singh2014,Song2014a,Weber2014,Schneider2014,Nigues2015,Cole2015,Tsioutsios2017,Guettinger2017,Tavernarakis2018}. This effort has often been paid off with the improvement of sensing capabilities and the measurement of unexpected phenomena.

Care has to be taken to avoid heating when improving the detection of the motion. The transduction of the motion is achieved by applying some input power to the resonator. In the case of nanotube resonators, the input power is usually related to the oscillating voltage applied across the nanotube~\cite{Sazonova2004} or the laser beam illuminating the nanotube~\cite{Stapfner2013,Tavernarakis2018}. The displacement sensitivity becomes better when increasing the input power. However, the input power has to be kept low enough to avoid electrical Joule heating and optical adsorption heating. Heating is especially prominent in tiny objects, such as nanotubes, because of their small heat capacity. Heating is detrimental, because it deteriorates the force and the mass sensitivity and increases the number of quanta of vibrational energy.

Here, we report on a novel detection method that allows us to measure the mechanical vibrations of nanotube resonators with an unprecedented sensitivity. The detection consists in measuring the electrical signal with a RLC resonator and a high electron mobility transistor (HEMT) amplifier cooled at liquid-helium temperature. In order to further improve the detection, we develop a new fabrication process to enhance the capacitive coupling between the ultraclean carbon nanotube and the gate electrode. This allows us to achieve $1.7~\mathrm{pm}/\sqrt{\mathrm{Hz}}$ displacement sensitivity  when the temperature of the measured eigenmode is $120$~mK. At higher vibration temperature, the resonator can be probed with larger input power, so that the sensitivity reaches $0.5~\mathrm{pm}/\sqrt{\mathrm{Hz}}$ at 300~mK.

We use a new fabrication process to grow ultraclean carbon nanotube resonators suspended over shallow trenches. Figure~1a shows a $\sim1.3~\mu$m long nanotube contacted electrically to two electrodes and separated from the local gate electrode by $\sim150$~nm. The electrodes made from platinum with a tungsten adhesion layer are evaporated on top of silicon dioxide grown by plasma-enhanced chemical vapour deposition. Nanotubes are grown by the `fast heating' chemical vapour deposition method in the last fabrication step~\cite{Huang2004}. This method consists in rapidly sliding the quartz tube through the oven under a flow of methane, so that the sample moves from a position outside of the oven to the center of the oven, whose temperature is $T_\mathrm{growth}=820$~ $^{\circ}$C. This growth process has two assets compared to the usual growth of nanotube resonators~\cite{Moser2014}. It allows us to suspend nanotubes over wide trenches. In addition, the electrodes are less prone to melt and change shape.

Mechanical vibrations are detected electrically using a RLC resonator and a HEMT amplifier cooled at liquid-helium temperature (Fig.~1b). Displacement modulation is transduced capacitively into current modulation by applying an input oscillating voltage $V_\mathrm{sd}^\mathrm{ac}$ across the nanotube~\cite{Sazonova2004,Moser2013,Moser2014}. The frequency $\omega_\mathrm{sd}/2\pi$ of the oscillating voltage is set to match $\omega_\mathrm{sd}=\omega_\mathrm{0} \pm \omega_\mathrm{RLC}$, where $\omega_\mathrm{0}/2\pi$ is the resonance frequency of the nanotube resonator and $\omega_\mathrm{RLC}/2\pi=1.25$~MHz the resonance frequency of the RLC resonator.   Driven vibrations are measured with the two-source method~\cite{Sazonova2004}. Thermal vibrations are measured by recording the current noise at $\sim \omega_\mathrm{RLC}$~\cite{Moser2013,Moser2014}. These current noise measurements are similar to those recently carried out on quantum electron devices~\cite{Jezouin2013,Bocquillon2013,Jullien2014}.


The RLC resonator and the HEMT amplifier \cite{Dong2014} allow us to reduce the current noise floor at $\sim \omega_\mathrm{RLC}$ down to $28~\mathrm{fA}/\sqrt{\mathrm{Hz}}$ below $\sim 100$~mK (Fig.~1c). The current noise floor is temperature dependent above $\sim 100$~mK because of the Johnson-Nyquist noise of the impedance of the RLC resonator. Below $\sim 100$~mK, the Johnson-Nyquist noise becomes vanishingly small. The noise floor is then given by the current noise ($6.2~\mathrm{fA}/\sqrt{\mathrm{Hz}}$) and the voltage noise ($0.16~\mathrm{nV}/\sqrt{\mathrm{Hz}}$) of the HEMT amplifier and the voltage noise of the room temperature amplifier. The gain of the HEMT amplifier is set at 5.6. The inductance of the circuit is given by the 66~$\mu$H inductance soldered onto a printed-circuit board (PCB). The 242~pF capacitance measured from the RLC resonance frequency comes from the capacitance of the radio-frequency cables and the low-pass filter VLFX-80 between the device and the HEMT. The 7.52~k$\Omega$ resistance obtained from the 87~kHz line-width of the RLC resonator is attributed to the 10~k$\Omega$ resistance soldered onto the PCB and the input impedance of the HEMT amplifier.

The lowest-lying flexural eigenmodes are identified by capacitively driving the resonator with an oscillating force and measuring the motion with the two-source method~\cite{Sazonova2004}. The dependence of the resonance frequency as a function of the static voltage $V_\mathrm{G}^\mathrm{dc}$ applied to the gate electrode demonstrates that the measured resonance is related to a mechanical eigenmode of the nanotube (Fig.~2a). The amplitude of the lowest-frequency resonance is much larger than that of the second detected resonance (Fig.~2b). We conclude that the detected eignmodes are polarized in the direction perpendicular to the surface of the gate electrode to a good approximation, and that eigenmodes polarized in the parallel direction cannot be detected.

The thermal vibrations are recorded at the base temperature of the dilution cryostat (Fig.~2c). We switch off the driving force, and the displacement noise is recorded with the method described in Refs.~\cite{Moser2013,Moser2014}.  The quality factor is $Q=$627,000 with the gate voltage set at $V_\mathrm{G}^\mathrm{dc}$=-0.21~V. We choose this gate voltage so that the electron transport is not in the Coulomb blockade regime. The $Q$-factor becomes lower at more negative gate voltages because of electrical losses~\cite{Song2012} and at positive gate voltages due to Coulomb blockade~\cite{Lassagne2009}. We measure the dependence of the variance of the displacement $\left\langle \delta z^2 \right\rangle$ on the cryostat temperature $T$ (Fig.~2d). The linear dependence is in agreement with the equipartition theorem $m \omega_\mathrm{0}^2 \left\langle \delta z^2 \right\rangle =k_\mathrm b T$, where $m$ is the effective mass of the resonator. We obtain $m=8.6$~ag from the slope, which is consistent with the mass expected for a $\sim1.3~\mu$m long nanotube. Below $\sim 120$~mK the eigenmode does not thermalize well with the cryostat (Fig.~2d). Measurements on a second nanotube resonator show that the eigenmode reaches $\sim 50$~mK (Supplementary Information). The origin of this poor thermalization at low temperature may be related to a non-thermal force noise, such as the electrostatic force noise related to the voltage noise in the device~\cite{Weber2016}.


The force sensitivity derived from the noise spectrum in Fig.~2c is $\sqrt{S_\mathrm{FF}}=4.3\pm2.9~\mathrm{zN}/\sqrt{\mathrm{Hz}}$. Because the thermal resonance is described by a Lorentzian line shape, the force sensitivity is quantified from the total displacement noise at resonance frequency using $S_\mathrm{FF} = S_\mathrm{zz}(\omega_\mathrm 0) /|\chi(\omega_\mathrm 0)|^2$ with the mechanical susceptibility $|\chi(\omega_\mathrm 0)|=Q/m \omega_\mathrm 0^2$. The force sensitivity is given to a large extent by the thermal force noise of the resonator $\sqrt{S_\mathrm{FF}^\mathrm{th}}=\sqrt{4k_\mathrm{b}Tm\omega_\mathrm 0/Q}=4.0~\mathrm{zN}/\sqrt{\mathrm{Hz}}$, which is the fundamental limit of the force sensitivity set by the fluctuation-dissipation theorem. The noise of the imprecision in the detection contributes to the force sensitivity by a low amount. The error bar in the estimation of the force sensitivity originates essentially from the uncertainty in the nanotube diameter and the separation between the nanotube and the gate electrode. Table~\ref{table:ta} shows that the force sensitivity measured in this work is better than what is reported with resonators micro-fabricated from bulk material~\cite{Teufel2009,Heritier2018,Reinhardt2016}, and resonators based on nanotube~\cite{Moser2013}, semiconducting nanowire~\cite{Lepinay2016,Rossi2016,Nichol2012}, graphene~\cite{Weber2016}, and levitating particles~\cite{Gieseler2013a}. In Ref.~\cite{Moser2014}, the reported thermal force noise is lower, but the cross-correlation noise measurement does not quantify the force noise due to the imprecision in the detection, so that the total force noise cannot be quantified.

We now look at how the displacement noise is affected by the input power related to the oscillating voltage $V_\mathrm{sd}^\mathrm{ac}$ (Figs.~3a-d). The variance of the displacement increases abruptly above $V_\mathrm{sd}^\mathrm{ac} \simeq 80~\mu$V when the cryostat is at base temperature (Fig.~3c). This indicates the rise of the thermal vibration amplitude due to Joule heating. By contrast, the variance of the displacement remains constant over the whole range of $V_\mathrm{sd}^\mathrm{ac}$ that we apply when the cryostat temperature is set at 300~mK (Fig.~3d).

Our detection scheme allows us to reach an excellent displacement sensitivity for input powers below the onset of Joule heating. The displacement sensitivity is given by the noise floor of the spectrum of thermal vibrations. The displacement sensitivity gets better when increasing $V_\mathrm{sd}^\mathrm{ac}$ (Figs.~3e,f). The displacement sensitivity at base temperature is $1.7~\mathrm{pm}/\sqrt{\mathrm{Hz}}$ at $V_\mathrm{sd}^\mathrm{ac} = 80~\mu$V before that Joule heating starts to increase the variance of the displacement. When the cryostat temperature is set at 300~mK, the displacement sensitivity is $0.5~\mathrm{pm}/\sqrt{\mathrm{Hz}}$ at the largest $V_\mathrm{sd}^\mathrm{ac}$ value that we apply. The corresponding signal-to-noise ratio in the spectrum of thermal vibrations is 17~dB (Fig.~3b).  The measured displacement sensitivity $S_\mathrm{zz}^\mathrm{imp}$ scales as $(1/V_\mathrm{sd}^\mathrm{ac})^2$ (Figs~3e,f), indicating that $S_\mathrm{zz}^\mathrm{imp}$ is limited by the noise of the detection circuit and not by the electron shot noise through the nanotube~\cite{Wang2017a}. The $Q$-factor in Fig.~3 is lower than that in Fig.~2c due to an unknown reason while cycling the cryostat through room temperature; the resonance frequency and the mass are not modified by the thermal cycling.

In conclusion, we report on a detection scheme of nanotube resonators with an unprecedented displacement sensitivity. It allows us to reach $4.3~\mathrm{zN}/\sqrt{\mathrm{Hz}}$ force sensitivity, which surpasses what has been achieved with mechanical resonators to date. This high force sensitivity is an important step towards detecting individual nuclear spins with nuclear magnetic resonance measurements~\cite{Degen2009,Mamin2007a}. The coupling between mechanical vibrations and spins can be achieved by applying a gradient of magnetic field that is generated with the current biased through the gate electrode in Fig.~1a~\cite{Nichol2012}. In this context, the new fabrication process of nanotube resonators presented in this Letter is useful for increasing the gradient of the magnetic field, since it reduces the separation between the current-carrying electrode and the nanotube down to $\sim 150$~nm. The device layout might also allow us to carry out magnetic resonance force microscopy (MRFM) measurements to image the location of individual nuclear spins adsorbed along the nanotube. Imaging can be done by periodically applying radio-frequency pulses though the current-carrying electrode~\cite{Nichol2013}. Moreover, the advances in fabrication and detection described in this work offer new possibilities for studying the strong coupling between electrons and vibrations in nanoscale resonators~\cite{Armour2004,Clerk2005,Naik2006,Micchi2015}. In the Coulomb blockade regime, the system has been predicted to feature a transition towards a mechanically bistable and blocked-current state~\cite{Pistolesi2007,Weick2011,Micchi2015,Micchi2016}. This hitherto unobserved transition is expected to occur at higher temperature for shorter separation between the nanotube and the gate electrode~\cite{Micchi2015}. The high quality displacement noise spectra reported here might allow us to study this transition in details~\cite{Micchi2015} as well as the different cooling schemes that have been proposed theoretically using different quantum electron transport regimes~\cite{Zippilli2009,Santandrea2011,Stadler2014,Arrachea2014,Stadler2016}.

\textbf{Associated content}  Further information on the calibration of the displacement sensitivity and the force sensitivity; mechanical and electrical characterization of the device; displacement variance as a function of temperature of a second device.

\textbf{Acknowledgments} This work is supported by the ERC advanced grant 692876, the Foundation Cellex, the CERCA Programme, AGAUR, Severo Ochoa (SEV-2015-0522), the grant FIS2015-69831-P of MINECO, and the Fondo Europeo de Desarrollo Regional (FEDER).

\bibliographystyle{apsrev4-1}
\bibliography{../../../adrian2}

\begin{figure*}[h]
	\includegraphics[width=10cm]{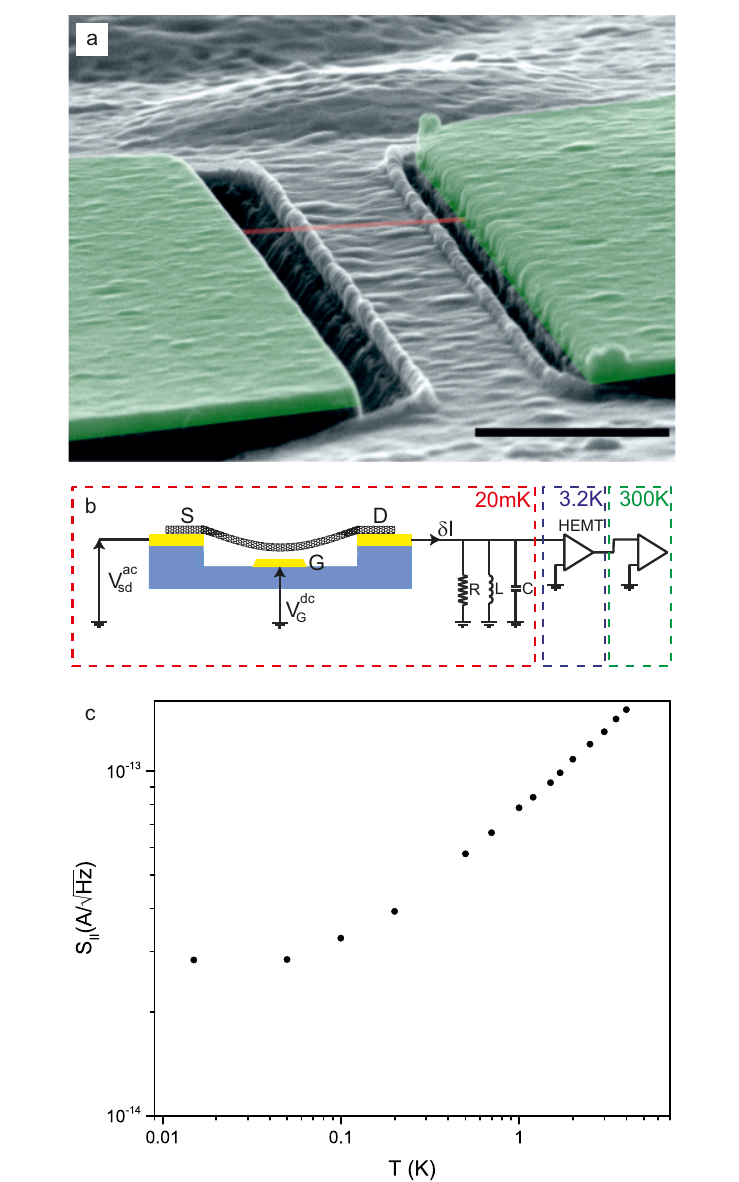}
	\caption{\textbf{Nantotube resonator and electrical circuit for the detection of the vibrations.} (\textbf{a}) False-colour scanning electron microscopy image of a typical nanotube resonator fabricated with the `fast heating' chemical vapour deposition method. The $\sim 20$~nm high ridges at the edges of the gate electrodes are attributed to resist residues. The scale bar is 1~$\mu$m. (\textbf{b}) Schematic of the measurement of the nanotube vibrations using the RLC resonator and the HEMT amplifier cooled at 3.2~K. The base temperature of the cryostat is $\sim 20$~mK. An oscillating voltage with amplitude $V_\mathrm{sd}^\mathrm{ac}$ is applied between electrodes $S$ and $D$, and a constant voltage $V_\mathrm{G}^\mathrm{dc}$ is applied to electrode $G$. (\textbf{c}) Temperature dependence of the current noise floor of the circuit measured at $\omega_\mathrm{RLC}$.
}
	\label{fig1}
\end{figure*}

\begin{figure*}[h]
	\includegraphics[width=12cm]{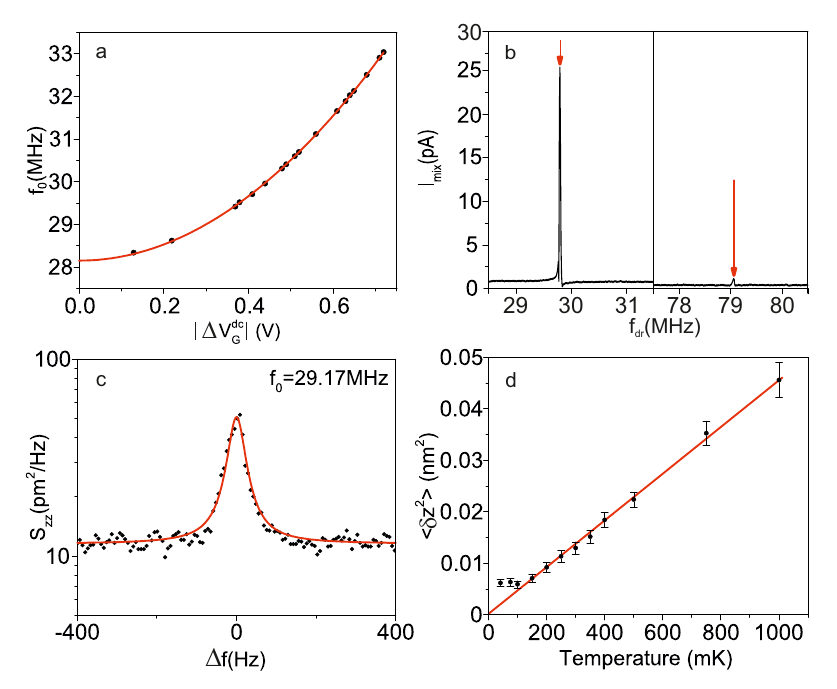}
	\caption{\textbf{Driven and thermal vibrations of the nanotube resonator.} (\textbf{a}) Gate voltage dependence of the resonance frequency of the fundamental eigenmode. The small positive offset voltage $V_\mathrm{off}=0.119$~V due to the work function difference between the nanotube and the gate electrode is subtracted from the applied $V_\mathrm{G}^\mathrm{dc}$ value. (\textbf{b}) Driven response of the two lowest-frequency detected mechanical eigenmodes as a function of the drive frequency measured with the two-source method.  The resonances are indicated by two red arrows. (\textbf{c}) Spectrum of the displacement noise of the fundamental eigenmode measured at the base temperature of the cryostat when applying $V_\mathrm{G}^\mathrm{dc}$=-0.21~V and $V_\mathrm{sd}^\mathrm{ac}=40~\mu$V. The resonance frequency $f_\mathrm{0}$ is given in the figure. (\textbf{d}) Variance of the displacement measured as a function of cryostat temperature.
}
	\label{fig2}
\end{figure*}

\begin{figure*}[h]
	\includegraphics[width=12cm]{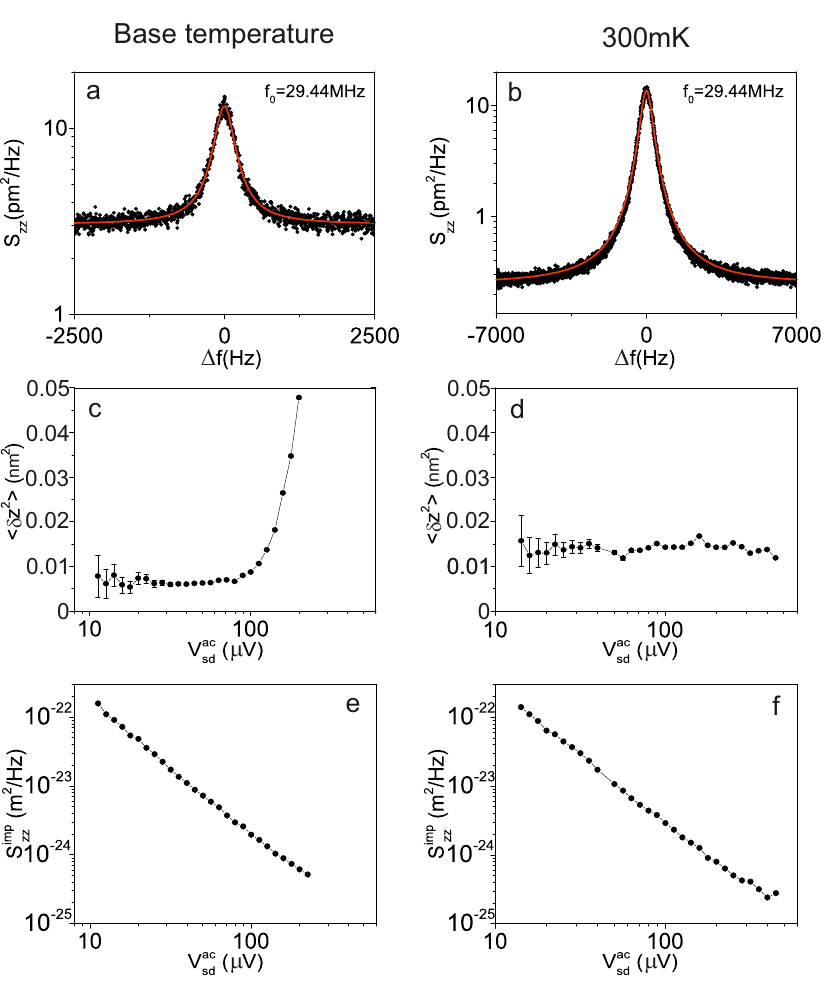}
	\caption{\textbf{Spectrum of the displacement noise modified by the oscillating voltage with amplitude $V_\mathrm{sd}^\mathrm{ac}$ applied across the nanotube.} (\textbf{a}) Spectrum of the displacement noise of the fundamental eigenmode measured at the base temperature of the cryostat when applying $V_\mathrm{G}^\mathrm{dc}$=-0.255~V and $V_\mathrm{sd}^\mathrm{ac}=70~\mu$V. (\textbf{b}) Same as \textbf{a} but with the cryostat temperature set at 300~mK and $V_\mathrm{sd}^\mathrm{ac}=400~\mu$V. (\textbf{c,d}) Dependence of the variance of the displacement on $V_\mathrm{sd}^\mathrm{ac}$ measured at the base temperature of the cryostat and 300~mK. (\textbf{e,f}) Dependence of the displacement sensitivity on $V_\mathrm{sd}^\mathrm{ac}$ measured at the base temperature of the cryostat and 300~mK.
}
	\label{fig3}
\end{figure*}

\begin{table*}[b]
\begin{center}
 \begin{tabular}{|p{3cm}|p{3cm}|p{3cm}|p{6cm}|}
\hline
 $\sqrt{S_\mathrm{FF}^\mathrm{th}}$ (N/$\sqrt{\mathrm{Hz}}$)& $\sqrt{S_\mathrm{FF}^\mathrm{imp}}$ (N/$\sqrt{\mathrm{Hz}}$) & $\sqrt{S_\mathrm{FF}}$ (N/$\sqrt{\mathrm{Hz}}$) & Description  \\
 \hline\hline
 $4.0\cdot 10^{-21}$  & $1.6\cdot 10^{-21}$ & $4.3\cdot 10^{-21}$ & Nanotube (this work) \\
 \hline
 $2.0\cdot 10^{-20}$  & negligible & $2.0\cdot 10^{-20}$ & levitating particle \cite{Gieseler2013a}\\
 \hline
 $2.7\cdot 10^{-19}$  & $2.7\cdot 10^{-19}$ & $3.9\cdot 10^{-19}$ & Graphene \cite{Weber2016}\\
 \hline
 $1.0\cdot 10^{-18}$  & negligible & $1.0\cdot 10^{-18}$ & Silicon nanowire \cite{Nichol2012}\\
  \hline
 $5.0\cdot 10^{-18}$  & negligible & $5.0\cdot 10^{-18}$ & GaAs/AlGaAs nanowire \cite{Rossi2016}\\
\hline
 $1.6\cdot 10^{-19}$  & $1.0\cdot 10^{-19}$ & $1.9\cdot 10^{-19}$ & Microfabricated ladder \cite{Heritier2018}\\
  \hline
 $5.1\cdot 10^{-19}$  & negligible & $5.1\cdot 10^{-19}$ & Microfabricated beam \cite{Teufel2009}\\
 \hline
 $2.0\cdot 10^{-17}$  & negligible & $2.0\cdot 10^{-17}$ & Microfabricated trampoline \cite{Reinhardt2016}\\
 \hline
 $1.2\cdot 10^{-20}$  & unknown & unknown & Nanotube \cite{Moser2013} \\
 \hline
 $\sim 1\cdot 10^{-21}$  & unknown & unknown & Nanotube \cite{Moser2014} \\
  \hline

\end{tabular}
\caption{
{\bf Thermal force noise $S_\mathrm{FF}^\mathrm{th}$, force noise due to the imprecision of the detection $S_\mathrm{FF}^\mathrm{imp}$, and total force sensitivity $S_\mathrm{FF}$ for different resonators. The three force noises are related by $S_\mathrm{FF}^\mathrm{th}+S_\mathrm{FF}^\mathrm{imp}=S_\mathrm{FF}$. }}
\label{table:ta}
\end{center}
\end{table*}

\newpage

\title{Supplementary Information\\Ultrasensitive displacement noise measurement of carbon nanotube mechanical resonators}

\author{S. de Bonis}
\author{C. Urgell}
\author{W. Yang}
\author{C. Samanta}
\author{A. Noury}
\author{J. Vergara-Cruz}
\author{Q. Dong}
\author{Y. Jin}
\author{A. Bachtold}

\maketitle


\section{Transduction of displacement into current}

We summarize the important relations for the transduction of displacement into current~\cite{Moser2014}. We use the fact that the mechanical eigenmode is polarized in the direction perpendicular to the surface of the gate electrode, as discussed in the main text. The current $\delta I$ at the frequency close to the difference between the
mode eigenfrequency and the frequency of the source-drain voltage
is
\begin{eqnarray}
&& \delta I=\beta\delta z ,
\label{eq:Iz}\\
&& \beta=\frac{1}{2}\frac{dG}{dV_\mathrm{G}}V_\mathrm{G}^\mathrm{dc}V_\mathrm{sd}^\mathrm{ac}\frac{C_\mathrm{G}^{\prime}}{C_\mathrm{G}}.
\label{eq:beta}
\end{eqnarray}
Here, $\delta z$ is the displacement of the nanotube, $dG/dV_\mathrm{G}$ is the transcondutance, $V_\mathrm{G}^\mathrm{dc}$ is the
static gate voltage, $V_\mathrm{sd}^\mathrm{ac}$ is the amplitude of the
oscillating source-drain voltage, $C_\mathrm{G}$ is the capacitance between the nanotube and the gate electrode, and $C_\mathrm{G}^{\prime}$ is the derivative of $C_\mathrm{G}$ with respect to $z$. We measure $dG/dV_\mathrm{G}=1.4\times10^{-3}$~S/V at the gate voltage discussed in the main text.

\begin{figure}[t]
\includegraphics{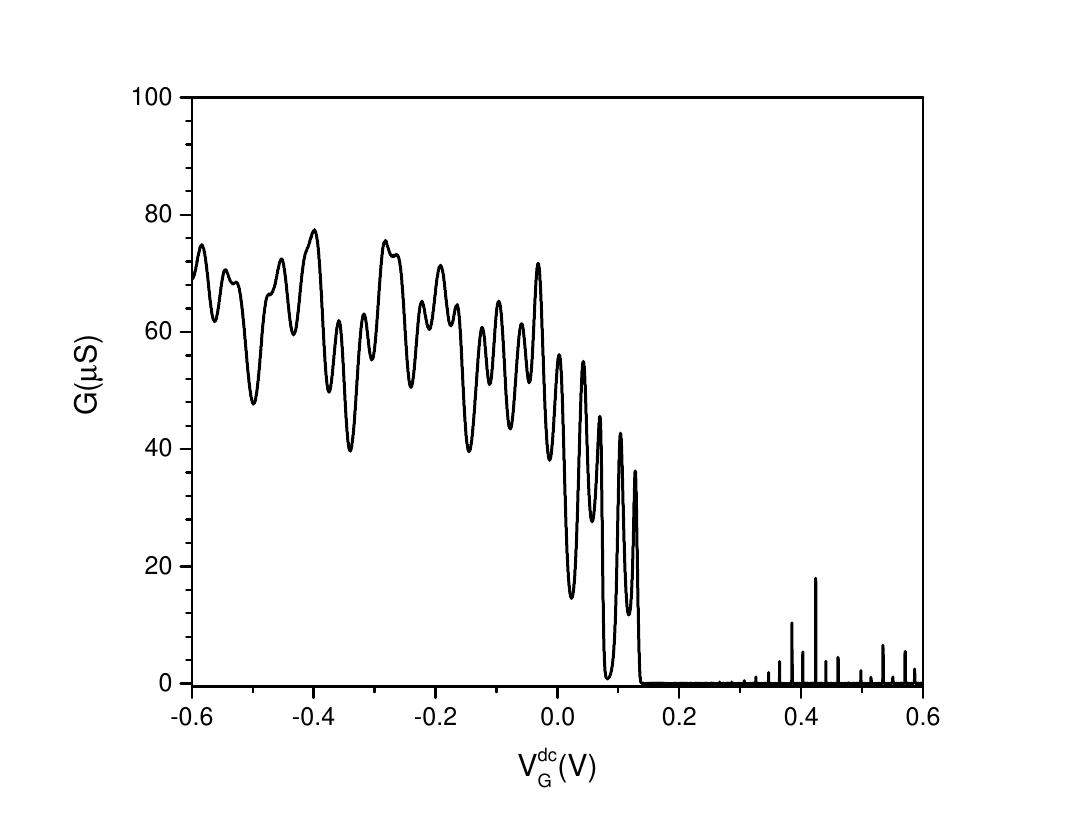}
\caption{Conductance of the nanotube as a function of gate voltage measured at the base temperature of the cryostat.}
\label{GVG}
\end{figure}

We estimate the capacitance $C_\mathrm{G}$ from the separation $\Delta V_\mathrm{G}^\mathrm{dc}=16.8\pm0.6$~mV between two
conductance peaks in the Coulomb blockade regime at large positive $V_\mathrm{G}^\mathrm{dc}$ values (Fig.~\ref{GVG}).
We obtain $C_\mathrm{G}=e/\Delta V_\mathrm{G}^\mathrm{dc}=9.5\pm0.3\times10^{-18}$~F. We quantify $C_\mathrm{G}^{\prime}$ using the relation
\begin{equation}
C_\mathrm{G}^{\prime}=\frac{C_\mathrm{G}}{d\ln(2d/r)}=\left(1.11\pm0.17\right)\times10^{-11}\textrm{F/m},
\label{Cprime}
\end{equation}
with $d=150\pm10$~nm for the separation between the nanotube and the gate electrode and $r=1\pm0.3$~nm for the radius of the nanotube.

The spectral density $S_\mathrm{zz}$ of the displacement noise in the main text is obtained from the measured spectral density of the current noise using Eqs.~\ref{eq:Iz} and \ref{eq:beta}. The displacement sensitivity $S_\mathrm{zz}^\mathrm{imp}$ in the main text is estimated from the current noise floor $S_\mathrm{II}^\mathrm{imp}$ in the measured spectrum using
\begin{equation}
S_\mathrm{zz}^\mathrm{imp}=\frac{1}{\beta^2}S_\mathrm{II}^\mathrm{imp}.
\label{Szzimp}
\end{equation}

\begin{figure}[t]
\includegraphics{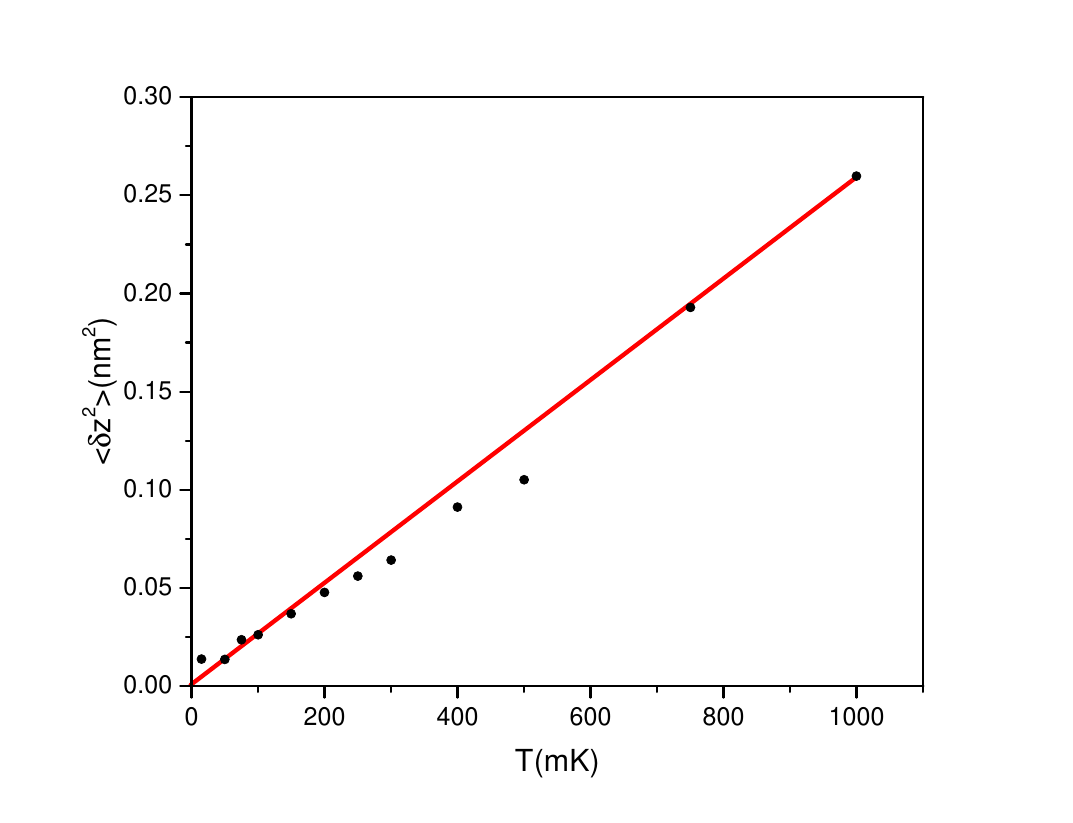}
\caption{Variance of the displacement of a second nanotube resonator as a function of the temperature of the cryostat.}
\label{resonator2}
\end{figure}

\section{Characteristics of the nanotube discussed in the main text}

We estimate the effective mass $m=8.6\pm3.6$~ag from the measurement of the variance of the displacement $\langle\delta z^{2}\rangle$ as a function of temperature $T$ in Fig.~2d of the main text. For this, we compare the measured slope of $\langle\delta z^{2}\rangle$ as a function of $T$ to the slope expected from the equipartition theorem, which reads
\begin{equation}
m\omega_{0}^2\langle\delta z^{2}\rangle=k_{b}T,
\label{equipartition}
\end{equation}
with $\omega_{0}/2\pi$ the resonance frequency of the eigenmode. This mass is consistent with the mass of a $\sim1.3~\mu$m long nanotube. We obtain the same mass through a second measurement, where we compare the amplitude of the driving vibration amplitude to the amplitude of the thermal vibrations, as discussed in the supplementary Information of Ref.~\cite{Moser2014}.

Because the thermal resonance is described by a Lorentzian line shape in Fig.~2c of the main text, the force sensitivity is quantified from the total displacement noise at resonance frequency using
\begin{equation}
\sqrt{S_\mathrm{FF}} = \sqrt{S_\mathrm{zz}(\omega_\mathrm 0)} \frac{m \omega_\mathrm 0^2}{Q}=\left(4.3\pm2.9\right)\times10^{-21}\textrm{N}/\sqrt{\mathrm{Hz}},
\label{SFF}
\end{equation}
where $Q$ is the quality factor measured from the linewidth in the spectrum. The error bar in the estimation of the force sensitivity originates essentially from the uncertainty in $d$ and $r$.

The electrical characteristics of the nanotube in Fig.~\ref{GVG} is typical of ultraclean nanotubes \cite{Moser2014}. For large positive $V_\mathrm{G}^\mathrm{dc}$ values, $p-n$ junctions are
formed near the metal electrodes, forming a Coulomb blockaded region along the suspended nanotube. For negative $V_\mathrm{G}^\mathrm{dc}$, the nanotube is $p$-doped along the whole tube, resulting in a larger conductance.

\section{Thermalization of a second nanotube resonator}

Figure~\ref{resonator2} shows the variance of the displacement as a function of temperature. We observe a linear dependence of $\langle\delta z^{2}\rangle$ as a function of $T$ at high temperature. The variance $\langle\delta z^{2}\rangle$ at the base temperature of the cryostat corresponds to a temperature of $\sim 50$~mK.

\bibliographystyle{apsrev4-1}
\bibliography{../../../adrian2}

\begin{thebibliography}{74}%
\makeatletter
\providecommand \@ifxundefined [1]{%
 \@ifx{#1\undefined}
}%
\providecommand \@ifnum [1]{%
 \ifnum #1\expandafter \@firstoftwo
 \else \expandafter \@secondoftwo
 \fi
}%
\providecommand \@ifx [1]{%
 \ifx #1\expandafter \@firstoftwo
 \else \expandafter \@secondoftwo
 \fi
}%
\providecommand \natexlab [1]{#1}%
\providecommand \enquote  [1]{``#1''}%
\providecommand \bibnamefont  [1]{#1}%
\providecommand \bibfnamefont [1]{#1}%
\providecommand \citenamefont [1]{#1}%
\providecommand \href@noop [0]{\@secondoftwo}%
\providecommand \href [0]{\begingroup \@sanitize@url \@href}%
\providecommand \@href[1]{\@@startlink{#1}\@@href}%
\providecommand \@@href[1]{\endgroup#1\@@endlink}%
\providecommand \@sanitize@url [0]{\catcode `\\12\catcode `\$12\catcode
  `\&12\catcode `\#12\catcode `\^12\catcode `\_12\catcode `\%12\relax}%
\providecommand \@@startlink[1]{}%
\providecommand \@@endlink[0]{}%
\providecommand \url  [0]{\begingroup\@sanitize@url \@url }%
\providecommand \@url [1]{\endgroup\@href {#1}{\urlprefix }}%
\providecommand \urlprefix  [0]{URL }%
\providecommand \Eprint [0]{\href }%
\providecommand \doibase [0]{http://dx.doi.org/}%
\providecommand \selectlanguage [0]{\@gobble}%
\providecommand \bibinfo  [0]{\@secondoftwo}%
\providecommand \bibfield  [0]{\@secondoftwo}%
\providecommand \translation [1]{[#1]}%
\providecommand \BibitemOpen [0]{}%
\providecommand \bibitemStop [0]{}%
\providecommand \bibitemNoStop [0]{.\EOS\space}%
\providecommand \EOS [0]{\spacefactor3000\relax}%
\providecommand \BibitemShut  [1]{\csname bibitem#1\endcsname}%
\let\auto@bib@innerbib\@empty
\bibitem [{\citenamefont {Sazonova}\ \emph {et~al.}(2004)\citenamefont
  {Sazonova}, \citenamefont {Yaish}, \citenamefont {Ustunel}, \citenamefont
  {Roundy}, \citenamefont {Arias},\ and\ \citenamefont
  {McEuen}}]{Sazonova2004}%
  \BibitemOpen
  \bibfield  {author} {\bibinfo {author} {\bibfnamefont {V.}~\bibnamefont
  {Sazonova}}, \bibinfo {author} {\bibfnamefont {Y.}~\bibnamefont {Yaish}},
  \bibinfo {author} {\bibfnamefont {H.}~\bibnamefont {Ustunel}}, \bibinfo
  {author} {\bibfnamefont {D.}~\bibnamefont {Roundy}}, \bibinfo {author}
  {\bibfnamefont {T.~A.}\ \bibnamefont {Arias}}, \ and\ \bibinfo {author}
  {\bibfnamefont {P.~L.}\ \bibnamefont {McEuen}},\ }\href@noop {} {\bibfield
  {journal} {\bibinfo  {journal} {Nature}\ }\textbf {\bibinfo {volume} {431}},\
  \bibinfo {pages} {284} (\bibinfo {year} {2004})}\BibitemShut {NoStop}%
\bibitem [{\citenamefont {Bunch}\ \emph {et~al.}(2007)\citenamefont {Bunch},
  \citenamefont {Zande}, \citenamefont {Verbridge}, \citenamefont {Frank},
  \citenamefont {Tanenbaum}, \citenamefont {Parpia}, \citenamefont
  {Craighead},\ and\ \citenamefont {McEuen}}]{Bunch2007}%
  \BibitemOpen
  \bibfield  {author} {\bibinfo {author} {\bibfnamefont {J.~S.}\ \bibnamefont
  {Bunch}}, \bibinfo {author} {\bibfnamefont {A.~M. v.~d.}\ \bibnamefont
  {Zande}}, \bibinfo {author} {\bibfnamefont {S.~S.}\ \bibnamefont
  {Verbridge}}, \bibinfo {author} {\bibfnamefont {I.~W.}\ \bibnamefont
  {Frank}}, \bibinfo {author} {\bibfnamefont {D.~M.}\ \bibnamefont
  {Tanenbaum}}, \bibinfo {author} {\bibfnamefont {J.~M.}\ \bibnamefont
  {Parpia}}, \bibinfo {author} {\bibfnamefont {H.~G.}\ \bibnamefont
  {Craighead}}, \ and\ \bibinfo {author} {\bibfnamefont {P.~L.}\ \bibnamefont
  {McEuen}},\ }\href@noop {} {\bibfield  {journal} {\bibinfo  {journal}
  {Science}\ }\textbf {\bibinfo {volume} {315}},\ \bibinfo {pages} {490}
  (\bibinfo {year} {2007})}\BibitemShut {NoStop}%
\bibitem [{\citenamefont {Singh}\ \emph {et~al.}(2010)\citenamefont {Singh},
  \citenamefont {Sengupta}, \citenamefont {Solanki}, \citenamefont {Dhall},
  \citenamefont {Allain}, \citenamefont {Dhara}, \citenamefont {Pant},\ and\
  \citenamefont {Deshmukh}}]{Singh2010}%
  \BibitemOpen
  \bibfield  {author} {\bibinfo {author} {\bibfnamefont {V.}~\bibnamefont
  {Singh}}, \bibinfo {author} {\bibfnamefont {S.}~\bibnamefont {Sengupta}},
  \bibinfo {author} {\bibfnamefont {H.~S.}\ \bibnamefont {Solanki}}, \bibinfo
  {author} {\bibfnamefont {R.}~\bibnamefont {Dhall}}, \bibinfo {author}
  {\bibfnamefont {A.}~\bibnamefont {Allain}}, \bibinfo {author} {\bibfnamefont
  {S.}~\bibnamefont {Dhara}}, \bibinfo {author} {\bibfnamefont
  {P.}~\bibnamefont {Pant}}, \ and\ \bibinfo {author} {\bibfnamefont {M.~M.}\
  \bibnamefont {Deshmukh}},\ }\href
  {http://stacks.iop.org/0957-4484/21/i=16/a=165204} {\bibfield  {journal}
  {\bibinfo  {journal} {Nanotechnology}\ }\textbf {\bibinfo {volume} {21}},\
  \bibinfo {pages} {165204} (\bibinfo {year} {2010})}\BibitemShut {NoStop}%
\bibitem [{\citenamefont {Will}\ \emph {et~al.}(2017)\citenamefont {Will},
  \citenamefont {Hamer}, \citenamefont {Muller}, \citenamefont {Noury},
  \citenamefont {Weber}, \citenamefont {Bachtold}, \citenamefont {Gorbachev},
  \citenamefont {Stampfer},\ and\ \citenamefont {Guttinger}}]{Will2017}%
  \BibitemOpen
  \bibfield  {author} {\bibinfo {author} {\bibfnamefont {M.}~\bibnamefont
  {Will}}, \bibinfo {author} {\bibfnamefont {M.}~\bibnamefont {Hamer}},
  \bibinfo {author} {\bibfnamefont {M.}~\bibnamefont {Muller}}, \bibinfo
  {author} {\bibfnamefont {A.}~\bibnamefont {Noury}}, \bibinfo {author}
  {\bibfnamefont {P.}~\bibnamefont {Weber}}, \bibinfo {author} {\bibfnamefont
  {A.}~\bibnamefont {Bachtold}}, \bibinfo {author} {\bibfnamefont {R.~V.}\
  \bibnamefont {Gorbachev}}, \bibinfo {author} {\bibfnamefont {C.}~\bibnamefont
  {Stampfer}}, \ and\ \bibinfo {author} {\bibfnamefont {J.}~\bibnamefont
  {Guttinger}},\ }\href {\doibase 10.1021/acs.nanolett.7b01845} {\bibfield
  {journal} {\bibinfo  {journal} {Nano Lett.}\ }\textbf {\bibinfo {volume}
  {17}},\ \bibinfo {pages} {5950} (\bibinfo {year} {2017})}\BibitemShut
  {NoStop}%
\bibitem [{\citenamefont {Feng}\ \emph {et~al.}(2007)\citenamefont {Feng},
  \citenamefont {He}, \citenamefont {Yang},\ and\ \citenamefont
  {Roukes}}]{Feng2007}%
  \BibitemOpen
  \bibfield  {author} {\bibinfo {author} {\bibfnamefont {X.~L.}\ \bibnamefont
  {Feng}}, \bibinfo {author} {\bibfnamefont {R.}~\bibnamefont {He}}, \bibinfo
  {author} {\bibfnamefont {P.}~\bibnamefont {Yang}}, \ and\ \bibinfo {author}
  {\bibfnamefont {M.~L.}\ \bibnamefont {Roukes}},\ }\href {\doibase
  10.1021/nl0706695} {\bibfield  {journal} {\bibinfo  {journal} {Nano Lett.}\
  }\textbf {\bibinfo {volume} {7}},\ \bibinfo {pages} {1953} (\bibinfo {year}
  {2007})}\BibitemShut {NoStop}%
\bibitem [{\citenamefont {Gil-Santos}\ \emph {et~al.}(2010)\citenamefont
  {Gil-Santos}, \citenamefont {Ramos}, \citenamefont {Martinez}, \citenamefont
  {Fernandez-Regulez}, \citenamefont {Garcia}, \citenamefont {San~Paulo},
  \citenamefont {Calleja},\ and\ \citenamefont {Tamayo}}]{Gil-Santos2010}%
  \BibitemOpen
  \bibfield  {author} {\bibinfo {author} {\bibfnamefont {E.}~\bibnamefont
  {Gil-Santos}}, \bibinfo {author} {\bibfnamefont {D.}~\bibnamefont {Ramos}},
  \bibinfo {author} {\bibfnamefont {J.}~\bibnamefont {Martinez}}, \bibinfo
  {author} {\bibfnamefont {M.}~\bibnamefont {Fernandez-Regulez}}, \bibinfo
  {author} {\bibfnamefont {R.}~\bibnamefont {Garcia}}, \bibinfo {author}
  {\bibfnamefont {A.}~\bibnamefont {San~Paulo}}, \bibinfo {author}
  {\bibfnamefont {M.}~\bibnamefont {Calleja}}, \ and\ \bibinfo {author}
  {\bibfnamefont {J.}~\bibnamefont {Tamayo}},\ }\href
  {http://dx.doi.org/10.1038/nnano.2010.151} {\bibfield  {journal} {\bibinfo
  {journal} {Nature Nanotechnology}\ }\textbf {\bibinfo {volume} {5}},\
  \bibinfo {pages} {641} (\bibinfo {year} {2010})}\BibitemShut {NoStop}%
\bibitem [{\citenamefont {Sansa}\ \emph {et~al.}(2012)\citenamefont {Sansa},
  \citenamefont {Fernandez-Regulez}, \citenamefont {San~Paulo},\ and\
  \citenamefont {Perez-Murano}}]{Sansa2012}%
  \BibitemOpen
  \bibfield  {author} {\bibinfo {author} {\bibfnamefont {M.}~\bibnamefont
  {Sansa}}, \bibinfo {author} {\bibfnamefont {M.}~\bibnamefont
  {Fernandez-Regulez}}, \bibinfo {author} {\bibfnamefont {A.}~\bibnamefont
  {San~Paulo}}, \ and\ \bibinfo {author} {\bibfnamefont {F.}~\bibnamefont
  {Perez-Murano}},\ }\href {\doibase 10.1063/1.4771982} {\bibfield  {journal}
  {\bibinfo  {journal} {Appl. Phys. Lett.}\ }\textbf {\bibinfo {volume}
  {101}},\ \bibinfo {pages} {243115} (\bibinfo {year} {2012})}\BibitemShut
  {NoStop}%
\bibitem [{\citenamefont {Montinaro}\ \emph {et~al.}(2014)\citenamefont
  {Montinaro}, \citenamefont {Wust}, \citenamefont {Munsch}, \citenamefont
  {Fontana}, \citenamefont {Russo-Averchi}, \citenamefont {Heiss},
  \citenamefont {Fontcuberta~i Morral}, \citenamefont {Warburton},\ and\
  \citenamefont {Poggio}}]{Montinaro2014}%
  \BibitemOpen
  \bibfield  {author} {\bibinfo {author} {\bibfnamefont {M.}~\bibnamefont
  {Montinaro}}, \bibinfo {author} {\bibfnamefont {G.}~\bibnamefont {Wust}},
  \bibinfo {author} {\bibfnamefont {M.}~\bibnamefont {Munsch}}, \bibinfo
  {author} {\bibfnamefont {Y.}~\bibnamefont {Fontana}}, \bibinfo {author}
  {\bibfnamefont {E.}~\bibnamefont {Russo-Averchi}}, \bibinfo {author}
  {\bibfnamefont {M.}~\bibnamefont {Heiss}}, \bibinfo {author} {\bibfnamefont
  {A.}~\bibnamefont {Fontcuberta~i Morral}}, \bibinfo {author} {\bibfnamefont
  {R.~J.}\ \bibnamefont {Warburton}}, \ and\ \bibinfo {author} {\bibfnamefont
  {M.}~\bibnamefont {Poggio}},\ }\href {\doibase 10.1021/nl501413t} {\bibfield
  {journal} {\bibinfo  {journal} {Nano Lett.}\ }\textbf {\bibinfo {volume}
  {14}},\ \bibinfo {pages} {4454} (\bibinfo {year} {2014})}\BibitemShut
  {NoStop}%
\bibitem [{\citenamefont {Gieseler}\ \emph {et~al.}(2012)\citenamefont
  {Gieseler}, \citenamefont {Deutsch}, \citenamefont {Quidant},\ and\
  \citenamefont {Novotny}}]{Gieseler2012}%
  \BibitemOpen
  \bibfield  {author} {\bibinfo {author} {\bibfnamefont {J.}~\bibnamefont
  {Gieseler}}, \bibinfo {author} {\bibfnamefont {B.}~\bibnamefont {Deutsch}},
  \bibinfo {author} {\bibfnamefont {R.}~\bibnamefont {Quidant}}, \ and\
  \bibinfo {author} {\bibfnamefont {L.}~\bibnamefont {Novotny}},\ }\href
  {https://link.aps.org/doi/10.1103/PhysRevLett.109.103603} {\bibfield
  {journal} {\bibinfo  {journal} {Phys. Rev. Lett.}\ }\textbf {\bibinfo
  {volume} {109}},\ \bibinfo {pages} {103603} (\bibinfo {year}
  {2012})}\BibitemShut {NoStop}%
\bibitem [{\citenamefont {Kiesel}\ \emph {et~al.}(2013)\citenamefont {Kiesel},
  \citenamefont {Blaser}, \citenamefont {Delic}, \citenamefont {Grass},
  \citenamefont {Kaltenbaek},\ and\ \citenamefont {Aspelmeyer}}]{Kiesel2013}%
  \BibitemOpen
  \bibfield  {author} {\bibinfo {author} {\bibfnamefont {N.}~\bibnamefont
  {Kiesel}}, \bibinfo {author} {\bibfnamefont {F.}~\bibnamefont {Blaser}},
  \bibinfo {author} {\bibfnamefont {U.}~\bibnamefont {Delic}}, \bibinfo
  {author} {\bibfnamefont {D.}~\bibnamefont {Grass}}, \bibinfo {author}
  {\bibfnamefont {R.}~\bibnamefont {Kaltenbaek}}, \ and\ \bibinfo {author}
  {\bibfnamefont {M.}~\bibnamefont {Aspelmeyer}},\ }\href
  {http://www.pnas.org/content/110/35/14180.abstract N2 - The coupling of a
  levitated submicron particle and an optical cavity field promises access to a
  unique parameter regime both for macroscopic quantum experiments and for
  high-precision force sensing. We report a demonstration of such controlled
  interactions by cavity cooling the center-of-mass motion of an optically
  trapped submicron particle. This paves the way for a light-matter interface
  that can enable room-temperature quantum experiments with mesoscopic
  mechanical systems.} {\bibfield  {journal} {\bibinfo  {journal} {Proceedings
  of the National Academy of Sciences}\ }\textbf {\bibinfo {volume} {110}},\
  \bibinfo {pages} {14180} (\bibinfo {year} {2013})}\BibitemShut {NoStop}%
\bibitem [{\citenamefont {Nichol}\ \emph {et~al.}(2008)\citenamefont {Nichol},
  \citenamefont {Hemesath}, \citenamefont {Lauhon},\ and\ \citenamefont
  {Budakian}}]{Nichol2008}%
  \BibitemOpen
  \bibfield  {author} {\bibinfo {author} {\bibfnamefont {J.~M.}\ \bibnamefont
  {Nichol}}, \bibinfo {author} {\bibfnamefont {E.~R.}\ \bibnamefont
  {Hemesath}}, \bibinfo {author} {\bibfnamefont {L.~J.}\ \bibnamefont
  {Lauhon}}, \ and\ \bibinfo {author} {\bibfnamefont {R.}~\bibnamefont
  {Budakian}},\ }\href {\doibase 10.1063/1.3025305} {\bibfield  {journal}
  {\bibinfo  {journal} {Appl. Phys. Lett.}\ }\textbf {\bibinfo {volume} {93}},\
  \bibinfo {pages} {193110} (\bibinfo {year} {2008})}\BibitemShut {NoStop}%
\bibitem [{\citenamefont {Moser}\ \emph {et~al.}(2013)\citenamefont {Moser},
  \citenamefont {G\"uttinger}, \citenamefont {Eichler}, \citenamefont
  {Esplandiu}, \citenamefont {Liu}, \citenamefont {Dykman},\ and\ \citenamefont
  {Bachtold}}]{Moser2013}%
  \BibitemOpen
  \bibfield  {author} {\bibinfo {author} {\bibfnamefont {J.}~\bibnamefont
  {Moser}}, \bibinfo {author} {\bibfnamefont {J.}~\bibnamefont {G\"uttinger}},
  \bibinfo {author} {\bibfnamefont {A.}~\bibnamefont {Eichler}}, \bibinfo
  {author} {\bibfnamefont {M.~J.}\ \bibnamefont {Esplandiu}}, \bibinfo {author}
  {\bibfnamefont {D.~E.}\ \bibnamefont {Liu}}, \bibinfo {author} {\bibfnamefont
  {M.~I.}\ \bibnamefont {Dykman}}, \ and\ \bibinfo {author} {\bibfnamefont
  {A.}~\bibnamefont {Bachtold}},\ }\href@noop {} {\bibfield  {journal}
  {\bibinfo  {journal} {Nat. Nanotech.}\ }\textbf {\bibinfo {volume} {8}},\
  \bibinfo {pages} {493} (\bibinfo {year} {2013})}\BibitemShut {NoStop}%
\bibitem [{\citenamefont {de~Lepinay}\ \emph {et~al.}(2016)\citenamefont
  {de~Lepinay}, \citenamefont {Pigeau}, \citenamefont {Besga}, \citenamefont
  {Vincent}, \citenamefont {Poncharal},\ and\ \citenamefont
  {Arcizet}}]{Lepinay2016}%
  \BibitemOpen
  \bibfield  {author} {\bibinfo {author} {\bibfnamefont {L.~M.}\ \bibnamefont
  {de~Lepinay}}, \bibinfo {author} {\bibfnamefont {B.}~\bibnamefont {Pigeau}},
  \bibinfo {author} {\bibfnamefont {B.}~\bibnamefont {Besga}}, \bibinfo
  {author} {\bibfnamefont {P.}~\bibnamefont {Vincent}}, \bibinfo {author}
  {\bibfnamefont {P.}~\bibnamefont {Poncharal}}, \ and\ \bibinfo {author}
  {\bibfnamefont {O.}~\bibnamefont {Arcizet}},\ }\href
  {http://dx.doi.org/10.1038/nnano.2016.193} {\bibfield  {journal} {\bibinfo
  {journal} {Nature Nanotechnology}\ }\textbf {\bibinfo {volume} {12}},\
  \bibinfo {pages} {156} (\bibinfo {year} {2016})}\BibitemShut {NoStop}%
\bibitem [{\citenamefont {Rossi}\ \emph {et~al.}(2016)\citenamefont {Rossi},
  \citenamefont {Braakman}, \citenamefont {Cadeddu}, \citenamefont {Vasyukov},
  \citenamefont {Tutuncuoglu}, \citenamefont {Fontcuberta~i Morral},\ and\
  \citenamefont {Poggio}}]{Rossi2016}%
  \BibitemOpen
  \bibfield  {author} {\bibinfo {author} {\bibfnamefont {N.}~\bibnamefont
  {Rossi}}, \bibinfo {author} {\bibfnamefont {F.~R.}\ \bibnamefont {Braakman}},
  \bibinfo {author} {\bibfnamefont {D.}~\bibnamefont {Cadeddu}}, \bibinfo
  {author} {\bibfnamefont {D.}~\bibnamefont {Vasyukov}}, \bibinfo {author}
  {\bibfnamefont {G.}~\bibnamefont {Tutuncuoglu}}, \bibinfo {author}
  {\bibfnamefont {A.}~\bibnamefont {Fontcuberta~i Morral}}, \ and\ \bibinfo
  {author} {\bibfnamefont {M.}~\bibnamefont {Poggio}},\ }\href
  {http://dx.doi.org/10.1038/nnano.2016.189} {\bibfield  {journal} {\bibinfo
  {journal} {Nature Nanotechnology}\ }\textbf {\bibinfo {volume} {12}},\
  \bibinfo {pages} {150} (\bibinfo {year} {2016})}\BibitemShut {NoStop}%
\bibitem [{\citenamefont {Chiu}\ \emph {et~al.}(2008)\citenamefont {Chiu},
  \citenamefont {Hung}, \citenamefont {Postma},\ and\ \citenamefont
  {Bockrath}}]{Chiu2008}%
  \BibitemOpen
  \bibfield  {author} {\bibinfo {author} {\bibfnamefont {H.-Y.}\ \bibnamefont
  {Chiu}}, \bibinfo {author} {\bibfnamefont {P.}~\bibnamefont {Hung}}, \bibinfo
  {author} {\bibfnamefont {H.~W.~C.}\ \bibnamefont {Postma}}, \ and\ \bibinfo
  {author} {\bibfnamefont {M.}~\bibnamefont {Bockrath}},\ }\href@noop {}
  {\bibfield  {journal} {\bibinfo  {journal} {Nano Lett.}\ }\textbf {\bibinfo
  {volume} {8}},\ \bibinfo {pages} {4342} (\bibinfo {year} {2008})}\BibitemShut
  {NoStop}%
\bibitem [{\citenamefont {Wang}\ \emph {et~al.}(2010)\citenamefont {Wang},
  \citenamefont {Wei}, \citenamefont {Morse}, \citenamefont {Dash},
  \citenamefont {Vilches},\ and\ \citenamefont {Cobden}}]{Wang2010}%
  \BibitemOpen
  \bibfield  {author} {\bibinfo {author} {\bibfnamefont {Z.}~\bibnamefont
  {Wang}}, \bibinfo {author} {\bibfnamefont {J.}~\bibnamefont {Wei}}, \bibinfo
  {author} {\bibfnamefont {P.}~\bibnamefont {Morse}}, \bibinfo {author}
  {\bibfnamefont {J.~G.}\ \bibnamefont {Dash}}, \bibinfo {author}
  {\bibfnamefont {O.~E.}\ \bibnamefont {Vilches}}, \ and\ \bibinfo {author}
  {\bibfnamefont {D.~H.}\ \bibnamefont {Cobden}},\ }\href {\doibase
  10.1126/science.1182507} {\bibfield  {journal} {\bibinfo  {journal}
  {Science}\ }\textbf {\bibinfo {volume} {327}},\ \bibinfo {pages} {552}
  (\bibinfo {year} {2010})}\BibitemShut {NoStop}%
\bibitem [{\citenamefont {Chaste}\ \emph {et~al.}(2012)\citenamefont {Chaste},
  \citenamefont {Eichler}, \citenamefont {Moser}, \citenamefont {Ceballos},
  \citenamefont {Rurali},\ and\ \citenamefont {Bachtold}}]{Chaste2012}%
  \BibitemOpen
  \bibfield  {author} {\bibinfo {author} {\bibfnamefont {J.}~\bibnamefont
  {Chaste}}, \bibinfo {author} {\bibfnamefont {A.}~\bibnamefont {Eichler}},
  \bibinfo {author} {\bibfnamefont {J.}~\bibnamefont {Moser}}, \bibinfo
  {author} {\bibfnamefont {G.}~\bibnamefont {Ceballos}}, \bibinfo {author}
  {\bibfnamefont {R.}~\bibnamefont {Rurali}}, \ and\ \bibinfo {author}
  {\bibfnamefont {A.}~\bibnamefont {Bachtold}},\ }\href@noop {} {\bibfield
  {journal} {\bibinfo  {journal} {Nat. Nano}\ }\textbf {\bibinfo {volume}
  {7}},\ \bibinfo {pages} {301} (\bibinfo {year} {2012})}\BibitemShut {NoStop}%
\bibitem [{\citenamefont {Gieseler}\ \emph {et~al.}(2013)\citenamefont
  {Gieseler}, \citenamefont {Novotny},\ and\ \citenamefont
  {Quidant}}]{Gieseler2013a}%
  \BibitemOpen
  \bibfield  {author} {\bibinfo {author} {\bibfnamefont {J.}~\bibnamefont
  {Gieseler}}, \bibinfo {author} {\bibfnamefont {L.}~\bibnamefont {Novotny}}, \
  and\ \bibinfo {author} {\bibfnamefont {R.}~\bibnamefont {Quidant}},\ }\href
  {http://dx.doi.org/10.1038/nphys2798} {\bibfield  {journal} {\bibinfo
  {journal} {Nature Physics}\ }\textbf {\bibinfo {volume} {9}},\ \bibinfo
  {pages} {806} (\bibinfo {year} {2013})}\BibitemShut {NoStop}%
\bibitem [{\citenamefont {{Zhang}}\ \emph {et~al.}(2014)\citenamefont
  {{Zhang}}, \citenamefont {{Moser}}, \citenamefont {G\"uttinger},
  \citenamefont {{Bachtold}},\ and\ \citenamefont {{Dykman}}}]{Zhang2014}%
  \BibitemOpen
  \bibfield  {author} {\bibinfo {author} {\bibfnamefont {Y.}~\bibnamefont
  {{Zhang}}}, \bibinfo {author} {\bibfnamefont {J.}~\bibnamefont {{Moser}}},
  \bibinfo {author} {\bibfnamefont {J.}~\bibnamefont {G\"uttinger}}, \bibinfo
  {author} {\bibfnamefont {A.}~\bibnamefont {{Bachtold}}}, \ and\ \bibinfo
  {author} {\bibfnamefont {M.~I.}\ \bibnamefont {{Dykman}}},\ }\href@noop {}
  {\bibfield  {journal} {\bibinfo  {journal} {Phys. Rev. Lett.}\ }\textbf
  {\bibinfo {volume} {113}},\ \bibinfo {pages} {255502} (\bibinfo {year}
  {2014})}\BibitemShut {NoStop}%
\bibitem [{\citenamefont {Miao}\ \emph {et~al.}(2014)\citenamefont {Miao},
  \citenamefont {Yeom}, \citenamefont {Wang}, \citenamefont {Standley},\ and\
  \citenamefont {Bockrath}}]{Miao2014}%
  \BibitemOpen
  \bibfield  {author} {\bibinfo {author} {\bibfnamefont {T.~F.}\ \bibnamefont
  {Miao}}, \bibinfo {author} {\bibfnamefont {S.}~\bibnamefont {Yeom}}, \bibinfo
  {author} {\bibfnamefont {P.}~\bibnamefont {Wang}}, \bibinfo {author}
  {\bibfnamefont {B.}~\bibnamefont {Standley}}, \ and\ \bibinfo {author}
  {\bibfnamefont {M.}~\bibnamefont {Bockrath}},\ }\href {\doibase
  10.1021/nl403936a} {\bibfield  {journal} {\bibinfo  {journal} {Nano Lett.}\
  }\textbf {\bibinfo {volume} {14}},\ \bibinfo {pages} {2982} (\bibinfo {year}
  {2014})}\BibitemShut {NoStop}%
\bibitem [{\citenamefont {Eichler}\ \emph {et~al.}(2011)\citenamefont
  {Eichler}, \citenamefont {Moser}, \citenamefont {Chaste}, \citenamefont
  {Zdrojek}, \citenamefont {Wilson-Rae},\ and\ \citenamefont
  {Bachtold}}]{Eichler2011a}%
  \BibitemOpen
  \bibfield  {author} {\bibinfo {author} {\bibfnamefont {A.}~\bibnamefont
  {Eichler}}, \bibinfo {author} {\bibfnamefont {J.}~\bibnamefont {Moser}},
  \bibinfo {author} {\bibfnamefont {J.}~\bibnamefont {Chaste}}, \bibinfo
  {author} {\bibfnamefont {M.}~\bibnamefont {Zdrojek}}, \bibinfo {author}
  {\bibfnamefont {I.}~\bibnamefont {Wilson-Rae}}, \ and\ \bibinfo {author}
  {\bibfnamefont {A.}~\bibnamefont {Bachtold}},\ }\href@noop {} {\bibfield
  {journal} {\bibinfo  {journal} {Nature Nanotech.}\ }\textbf {\bibinfo
  {volume} {6}},\ \bibinfo {pages} {339} (\bibinfo {year} {2011})}\BibitemShut
  {NoStop}%
\bibitem [{\citenamefont {Deng}\ \emph {et~al.}(2016)\citenamefont {Deng},
  \citenamefont {Zhu}, \citenamefont {Wang}, \citenamefont {Zou}, \citenamefont
  {Wang}, \citenamefont {Li}, \citenamefont {Cao}, \citenamefont {Liu},
  \citenamefont {Li}, \citenamefont {Xiao}, \citenamefont {Guo}, \citenamefont
  {Jiang}, \citenamefont {Dai},\ and\ \citenamefont {Guo}}]{Deng2016}%
  \BibitemOpen
  \bibfield  {author} {\bibinfo {author} {\bibfnamefont {G.-W.}\ \bibnamefont
  {Deng}}, \bibinfo {author} {\bibfnamefont {D.}~\bibnamefont {Zhu}}, \bibinfo
  {author} {\bibfnamefont {X.-H.}\ \bibnamefont {Wang}}, \bibinfo {author}
  {\bibfnamefont {C.-L.}\ \bibnamefont {Zou}}, \bibinfo {author} {\bibfnamefont
  {J.-T.}\ \bibnamefont {Wang}}, \bibinfo {author} {\bibfnamefont {H.-O.}\
  \bibnamefont {Li}}, \bibinfo {author} {\bibfnamefont {G.}~\bibnamefont
  {Cao}}, \bibinfo {author} {\bibfnamefont {D.}~\bibnamefont {Liu}}, \bibinfo
  {author} {\bibfnamefont {Y.}~\bibnamefont {Li}}, \bibinfo {author}
  {\bibfnamefont {M.}~\bibnamefont {Xiao}}, \bibinfo {author} {\bibfnamefont
  {G.-C.}\ \bibnamefont {Guo}}, \bibinfo {author} {\bibfnamefont {K.-L.}\
  \bibnamefont {Jiang}}, \bibinfo {author} {\bibfnamefont {X.-C.}\ \bibnamefont
  {Dai}}, \ and\ \bibinfo {author} {\bibfnamefont {G.-P.}\ \bibnamefont
  {Guo}},\ }\href {\doibase 10.1021/acs.nanolett.6b01875} {\bibfield  {journal}
  {\bibinfo  {journal} {Nano Lett.}\ }\textbf {\bibinfo {volume} {16}},\
  \bibinfo {pages} {5456} (\bibinfo {year} {2016})}\BibitemShut {NoStop}%
\bibitem [{\citenamefont {De~Alba}\ \emph {et~al.}(2016)\citenamefont
  {De~Alba}, \citenamefont {Massel}, \citenamefont {Storch}, \citenamefont
  {Abhilash}, \citenamefont {Hui}, \citenamefont {McEuen}, \citenamefont
  {Craighead},\ and\ \citenamefont {Parpia}}]{Alba2016}%
  \BibitemOpen
  \bibfield  {author} {\bibinfo {author} {\bibfnamefont {R.}~\bibnamefont
  {De~Alba}}, \bibinfo {author} {\bibfnamefont {F.}~\bibnamefont {Massel}},
  \bibinfo {author} {\bibfnamefont {I.~R.}\ \bibnamefont {Storch}}, \bibinfo
  {author} {\bibfnamefont {T.~S.}\ \bibnamefont {Abhilash}}, \bibinfo {author}
  {\bibfnamefont {A.}~\bibnamefont {Hui}}, \bibinfo {author} {\bibfnamefont
  {P.~L.}\ \bibnamefont {McEuen}}, \bibinfo {author} {\bibfnamefont {H.~G.}\
  \bibnamefont {Craighead}}, \ and\ \bibinfo {author} {\bibfnamefont {J.~M.}\
  \bibnamefont {Parpia}},\ }\href@noop {} {\bibfield  {journal} {\bibinfo
  {journal} {Nat. Nano.}\ }\textbf {\bibinfo {volume} {11}},\ \bibinfo {pages}
  {741} (\bibinfo {year} {2016})}\BibitemShut {NoStop}%
\bibitem [{\citenamefont {Mathew}\ \emph {et~al.}(2016)\citenamefont {Mathew},
  \citenamefont {Patel}, \citenamefont {Borah}, \citenamefont {Vijay},\ and\
  \citenamefont {Deshmukh}}]{Mathew2016}%
  \BibitemOpen
  \bibfield  {author} {\bibinfo {author} {\bibfnamefont {J.~P.}\ \bibnamefont
  {Mathew}}, \bibinfo {author} {\bibfnamefont {R.~N.}\ \bibnamefont {Patel}},
  \bibinfo {author} {\bibfnamefont {A.}~\bibnamefont {Borah}}, \bibinfo
  {author} {\bibfnamefont {R.}~\bibnamefont {Vijay}}, \ and\ \bibinfo {author}
  {\bibfnamefont {M.~M.}\ \bibnamefont {Deshmukh}},\ }\href@noop {} {\bibfield
  {journal} {\bibinfo  {journal} {Nat Nano}\ }\textbf {\bibinfo {volume}
  {11}},\ \bibinfo {pages} {747} (\bibinfo {year} {2016})}\BibitemShut
  {NoStop}%
\bibitem [{\citenamefont {Lassagne}\ \emph {et~al.}(2009)\citenamefont
  {Lassagne}, \citenamefont {Tarakanov}, \citenamefont {Kinaret}, \citenamefont
  {Garcia-Sanchez},\ and\ \citenamefont {Bachtold}}]{Lassagne2009}%
  \BibitemOpen
  \bibfield  {author} {\bibinfo {author} {\bibfnamefont {B.}~\bibnamefont
  {Lassagne}}, \bibinfo {author} {\bibfnamefont {Y.}~\bibnamefont {Tarakanov}},
  \bibinfo {author} {\bibfnamefont {J.}~\bibnamefont {Kinaret}}, \bibinfo
  {author} {\bibfnamefont {D.}~\bibnamefont {Garcia-Sanchez}}, \ and\ \bibinfo
  {author} {\bibfnamefont {A.}~\bibnamefont {Bachtold}},\ }\href@noop {}
  {\bibfield  {journal} {\bibinfo  {journal} {Science}\ }\textbf {\bibinfo
  {volume} {325}},\ \bibinfo {pages} {1107} (\bibinfo {year}
  {2009})}\BibitemShut {NoStop}%
\bibitem [{\citenamefont {Steele}\ \emph {et~al.}(2009)\citenamefont {Steele},
  \citenamefont {Huttel}, \citenamefont {Witkamp}, \citenamefont {Poot},
  \citenamefont {Meerwaldt}, \citenamefont {Kouwenhoven},\ and\ \citenamefont
  {van~der Zant}}]{Steele2009}%
  \BibitemOpen
  \bibfield  {author} {\bibinfo {author} {\bibfnamefont {G.~A.}\ \bibnamefont
  {Steele}}, \bibinfo {author} {\bibfnamefont {A.~K.}\ \bibnamefont {Huttel}},
  \bibinfo {author} {\bibfnamefont {B.}~\bibnamefont {Witkamp}}, \bibinfo
  {author} {\bibfnamefont {M.}~\bibnamefont {Poot}}, \bibinfo {author}
  {\bibfnamefont {H.~B.}\ \bibnamefont {Meerwaldt}}, \bibinfo {author}
  {\bibfnamefont {L.~P.}\ \bibnamefont {Kouwenhoven}}, \ and\ \bibinfo {author}
  {\bibfnamefont {H.~S.~J.}\ \bibnamefont {van~der Zant}},\ }\href@noop {}
  {\bibfield  {journal} {\bibinfo  {journal} {Science}\ }\textbf {\bibinfo
  {volume} {325}},\ \bibinfo {pages} {1103} (\bibinfo {year}
  {2009})}\BibitemShut {NoStop}%
\bibitem [{\citenamefont {Ganzhorn}\ and\ \citenamefont
  {Wernsdorfer}(2012)}]{Ganzhorn2012}%
  \BibitemOpen
  \bibfield  {author} {\bibinfo {author} {\bibfnamefont {M.}~\bibnamefont
  {Ganzhorn}}\ and\ \bibinfo {author} {\bibfnamefont {W.}~\bibnamefont
  {Wernsdorfer}},\ }\href
  {https://link.aps.org/doi/10.1103/PhysRevLett.108.175502} {\bibfield
  {journal} {\bibinfo  {journal} {Phys. Rev. Lett.}\ }\textbf {\bibinfo
  {volume} {108}},\ \bibinfo {pages} {175502} (\bibinfo {year}
  {2012})}\BibitemShut {NoStop}%
\bibitem [{\citenamefont {Benyamini}\ \emph {et~al.}(2014)\citenamefont
  {Benyamini}, \citenamefont {Hamo}, \citenamefont {Kusminskiy}, \citenamefont
  {von Oppen},\ and\ \citenamefont {Ilani}}]{Benyamini2014}%
  \BibitemOpen
  \bibfield  {author} {\bibinfo {author} {\bibfnamefont {A.}~\bibnamefont
  {Benyamini}}, \bibinfo {author} {\bibfnamefont {A.}~\bibnamefont {Hamo}},
  \bibinfo {author} {\bibfnamefont {S.~V.}\ \bibnamefont {Kusminskiy}},
  \bibinfo {author} {\bibfnamefont {F.}~\bibnamefont {von Oppen}}, \ and\
  \bibinfo {author} {\bibfnamefont {S.}~\bibnamefont {Ilani}},\ }\href
  {http://dx.doi.org/10.1038/nphys2842} {\bibfield  {journal} {\bibinfo
  {journal} {Nature Physics}\ }\textbf {\bibinfo {volume} {10}},\ \bibinfo
  {pages} {151} (\bibinfo {year} {2014})}\BibitemShut {NoStop}%
\bibitem [{\citenamefont {Ares}\ \emph {et~al.}(2016)\citenamefont {Ares},
  \citenamefont {Pei}, \citenamefont {Mavalankar}, \citenamefont
  {Mergenthaler}, \citenamefont {Warner}, \citenamefont {Briggs},\ and\
  \citenamefont {Laird}}]{Ares2016}%
  \BibitemOpen
  \bibfield  {author} {\bibinfo {author} {\bibfnamefont {N.}~\bibnamefont
  {Ares}}, \bibinfo {author} {\bibfnamefont {T.}~\bibnamefont {Pei}}, \bibinfo
  {author} {\bibfnamefont {A.}~\bibnamefont {Mavalankar}}, \bibinfo {author}
  {\bibfnamefont {M.}~\bibnamefont {Mergenthaler}}, \bibinfo {author}
  {\bibfnamefont {J.~H.}\ \bibnamefont {Warner}}, \bibinfo {author}
  {\bibfnamefont {G.~A.~D.}\ \bibnamefont {Briggs}}, \ and\ \bibinfo {author}
  {\bibfnamefont {E.~A.}\ \bibnamefont {Laird}},\ }\href
  {https://link.aps.org/doi/10.1103/PhysRevLett.117.170801} {\bibfield
  {journal} {\bibinfo  {journal} {Phys. Rev. Lett.}\ }\textbf {\bibinfo
  {volume} {117}},\ \bibinfo {pages} {170801} (\bibinfo {year}
  {2016})}\BibitemShut {NoStop}%
\bibitem [{\citenamefont {Gloppe}\ \emph {et~al.}(2014)\citenamefont {Gloppe},
  \citenamefont {Verlot}, \citenamefont {Dupont-Ferrier}, \citenamefont
  {Siria}, \citenamefont {Poncharal}, \citenamefont {Bachelier}, \citenamefont
  {Vincent},\ and\ \citenamefont {Arcizet}}]{Gloppe2014}%
  \BibitemOpen
  \bibfield  {author} {\bibinfo {author} {\bibfnamefont {A.}~\bibnamefont
  {Gloppe}}, \bibinfo {author} {\bibfnamefont {P.}~\bibnamefont {Verlot}},
  \bibinfo {author} {\bibfnamefont {E.}~\bibnamefont {Dupont-Ferrier}},
  \bibinfo {author} {\bibfnamefont {A.}~\bibnamefont {Siria}}, \bibinfo
  {author} {\bibfnamefont {P.}~\bibnamefont {Poncharal}}, \bibinfo {author}
  {\bibfnamefont {G.}~\bibnamefont {Bachelier}}, \bibinfo {author}
  {\bibfnamefont {P.}~\bibnamefont {Vincent}}, \ and\ \bibinfo {author}
  {\bibfnamefont {O.}~\bibnamefont {Arcizet}},\ }\href
  {http://dx.doi.org/10.1038/nnano.2014.189} {\bibfield  {journal} {\bibinfo
  {journal} {Nature Nanotechnology}\ }\textbf {\bibinfo {volume} {9}},\
  \bibinfo {pages} {920} (\bibinfo {year} {2014})}\BibitemShut {NoStop}%
\bibitem [{\citenamefont {Reserbat-Plantey}\ \emph {et~al.}(2016)\citenamefont
  {Reserbat-Plantey}, \citenamefont {Schadler}, \citenamefont {Gaudreau},
  \citenamefont {Navickaite}, \citenamefont {Guttinger}, \citenamefont {Chang},
  \citenamefont {Toninelli}, \citenamefont {Bachtold},\ and\ \citenamefont
  {Koppens}}]{Reserbat-Plantey2016}%
  \BibitemOpen
  \bibfield  {author} {\bibinfo {author} {\bibfnamefont {A.}~\bibnamefont
  {Reserbat-Plantey}}, \bibinfo {author} {\bibfnamefont {K.~G.}\ \bibnamefont
  {Schadler}}, \bibinfo {author} {\bibfnamefont {L.}~\bibnamefont {Gaudreau}},
  \bibinfo {author} {\bibfnamefont {G.}~\bibnamefont {Navickaite}}, \bibinfo
  {author} {\bibfnamefont {J.}~\bibnamefont {Guttinger}}, \bibinfo {author}
  {\bibfnamefont {D.}~\bibnamefont {Chang}}, \bibinfo {author} {\bibfnamefont
  {C.}~\bibnamefont {Toninelli}}, \bibinfo {author} {\bibfnamefont
  {A.}~\bibnamefont {Bachtold}}, \ and\ \bibinfo {author} {\bibfnamefont
  {F.~H.~L.}\ \bibnamefont {Koppens}},\ }\href
  {http://dx.doi.org/10.1038/ncomms10218} {\bibfield  {journal} {\bibinfo
  {journal} {Nature Communications}\ }\textbf {\bibinfo {volume} {7}},\
  \bibinfo {pages} {10218} (\bibinfo {year} {2016})}\BibitemShut {NoStop}%
\bibitem [{\citenamefont {Purcell}\ \emph {et~al.}(2002)\citenamefont
  {Purcell}, \citenamefont {Vincent}, \citenamefont {Journet},\ and\
  \citenamefont {Binh}}]{Purcell2002}%
  \BibitemOpen
  \bibfield  {author} {\bibinfo {author} {\bibfnamefont {S.~T.}\ \bibnamefont
  {Purcell}}, \bibinfo {author} {\bibfnamefont {P.}~\bibnamefont {Vincent}},
  \bibinfo {author} {\bibfnamefont {C.}~\bibnamefont {Journet}}, \ and\
  \bibinfo {author} {\bibfnamefont {V.~T.}\ \bibnamefont {Binh}},\ }\href
  {https://link.aps.org/doi/10.1103/PhysRevLett.89.276103} {\bibfield
  {journal} {\bibinfo  {journal} {Phys. Rev. Lett.}\ }\textbf {\bibinfo
  {volume} {89}},\ \bibinfo {pages} {276103} (\bibinfo {year}
  {2002})}\BibitemShut {NoStop}%
\bibitem [{\citenamefont {Chen}\ \emph {et~al.}(2009)\citenamefont {Chen},
  \citenamefont {Rosenblatt}, \citenamefont {Bolotin}, \citenamefont {Kalb},
  \citenamefont {Kim}, \citenamefont {Kymissis}, \citenamefont {Stormer},
  \citenamefont {Heinz},\ and\ \citenamefont {Hone}}]{Chen2009}%
  \BibitemOpen
  \bibfield  {author} {\bibinfo {author} {\bibfnamefont {C.}~\bibnamefont
  {Chen}}, \bibinfo {author} {\bibfnamefont {S.}~\bibnamefont {Rosenblatt}},
  \bibinfo {author} {\bibfnamefont {K.~I.}\ \bibnamefont {Bolotin}}, \bibinfo
  {author} {\bibfnamefont {W.}~\bibnamefont {Kalb}}, \bibinfo {author}
  {\bibfnamefont {P.}~\bibnamefont {Kim}}, \bibinfo {author} {\bibfnamefont
  {I.}~\bibnamefont {Kymissis}}, \bibinfo {author} {\bibfnamefont {H.~L.}\
  \bibnamefont {Stormer}}, \bibinfo {author} {\bibfnamefont {T.~F.}\
  \bibnamefont {Heinz}}, \ and\ \bibinfo {author} {\bibfnamefont
  {J.}~\bibnamefont {Hone}},\ }\href {http://dx.doi.org/10.1038/nnano.2009.267}
  {\bibfield  {journal} {\bibinfo  {journal} {Nature Nanotechnology}\ }\textbf
  {\bibinfo {volume} {4}},\ \bibinfo {pages} {861} (\bibinfo {year}
  {2009})}\BibitemShut {NoStop}%
\bibitem [{\citenamefont {Huttel}\ \emph {et~al.}(2009)\citenamefont {Huttel},
  \citenamefont {Steele}, \citenamefont {Witkamp}, \citenamefont {Poot},
  \citenamefont {Kouwenhoven},\ and\ \citenamefont {van~der
  Zant}}]{Huttel2009}%
  \BibitemOpen
  \bibfield  {author} {\bibinfo {author} {\bibfnamefont {A.~K.}\ \bibnamefont
  {Huttel}}, \bibinfo {author} {\bibfnamefont {G.~A.}\ \bibnamefont {Steele}},
  \bibinfo {author} {\bibfnamefont {B.}~\bibnamefont {Witkamp}}, \bibinfo
  {author} {\bibfnamefont {M.}~\bibnamefont {Poot}}, \bibinfo {author}
  {\bibfnamefont {L.~P.}\ \bibnamefont {Kouwenhoven}}, \ and\ \bibinfo {author}
  {\bibfnamefont {H.~S.~J.}\ \bibnamefont {van~der Zant}},\ }\href@noop {}
  {\bibfield  {journal} {\bibinfo  {journal} {Nano Lett.}\ }\textbf {\bibinfo
  {volume} {9}},\ \bibinfo {pages} {2547} (\bibinfo {year} {2009})}\BibitemShut
  {NoStop}%
\bibitem [{\citenamefont {Gouttenoire}\ \emph {et~al.}(2010)\citenamefont
  {Gouttenoire}, \citenamefont {Barois}, \citenamefont {Perisanu},
  \citenamefont {Leclercq}, \citenamefont {Purcell}, \citenamefont {Vincent},\
  and\ \citenamefont {Ayari}}]{Gouttenoire2010}%
  \BibitemOpen
  \bibfield  {author} {\bibinfo {author} {\bibfnamefont {V.}~\bibnamefont
  {Gouttenoire}}, \bibinfo {author} {\bibfnamefont {T.}~\bibnamefont {Barois}},
  \bibinfo {author} {\bibfnamefont {S.}~\bibnamefont {Perisanu}}, \bibinfo
  {author} {\bibfnamefont {J.~L.}\ \bibnamefont {Leclercq}}, \bibinfo {author}
  {\bibfnamefont {S.~T.}\ \bibnamefont {Purcell}}, \bibinfo {author}
  {\bibfnamefont {P.}~\bibnamefont {Vincent}}, \ and\ \bibinfo {author}
  {\bibfnamefont {A.}~\bibnamefont {Ayari}},\ }\href@noop {} {\bibfield
  {journal} {\bibinfo  {journal} {Small}\ }\textbf {\bibinfo {volume} {6}},\
  \bibinfo {pages} {1060} (\bibinfo {year} {2010})}\BibitemShut {NoStop}%
\bibitem [{\citenamefont {Arcizet}\ \emph {et~al.}(2011)\citenamefont
  {Arcizet}, \citenamefont {Jacques}, \citenamefont {Siria}, \citenamefont
  {Poncharal}, \citenamefont {Vincent},\ and\ \citenamefont
  {Seidelin}}]{Arcizet2011}%
  \BibitemOpen
  \bibfield  {author} {\bibinfo {author} {\bibfnamefont {O.}~\bibnamefont
  {Arcizet}}, \bibinfo {author} {\bibfnamefont {V.}~\bibnamefont {Jacques}},
  \bibinfo {author} {\bibfnamefont {A.}~\bibnamefont {Siria}}, \bibinfo
  {author} {\bibfnamefont {P.}~\bibnamefont {Poncharal}}, \bibinfo {author}
  {\bibfnamefont {P.}~\bibnamefont {Vincent}}, \ and\ \bibinfo {author}
  {\bibfnamefont {S.}~\bibnamefont {Seidelin}},\ }\href@noop {} {\bibfield
  {journal} {\bibinfo  {journal} {Nat. Phys.}\ }\textbf {\bibinfo {volume}
  {7}},\ \bibinfo {pages} {879} (\bibinfo {year} {2011})}\BibitemShut {NoStop}%
\bibitem [{\citenamefont {Moser}\ \emph {et~al.}(2014)\citenamefont {Moser},
  \citenamefont {Eichler}, \citenamefont {Guttinger}, \citenamefont {Dykman},\
  and\ \citenamefont {Bachtold}}]{Moser2014}%
  \BibitemOpen
  \bibfield  {author} {\bibinfo {author} {\bibfnamefont {J.}~\bibnamefont
  {Moser}}, \bibinfo {author} {\bibfnamefont {A.}~\bibnamefont {Eichler}},
  \bibinfo {author} {\bibfnamefont {J.}~\bibnamefont {Guttinger}}, \bibinfo
  {author} {\bibfnamefont {M.~I.}\ \bibnamefont {Dykman}}, \ and\ \bibinfo
  {author} {\bibfnamefont {A.}~\bibnamefont {Bachtold}},\ }\href
  {http://dx.doi.org/10.1038/nnano.2014.234} {\bibfield  {journal} {\bibinfo
  {journal} {Nature Nanotechnology}\ }\textbf {\bibinfo {volume} {9}},\
  \bibinfo {pages} {1007} (\bibinfo {year} {2014})}\BibitemShut {NoStop}%
\bibitem [{\citenamefont {Stapfner}\ \emph {et~al.}(2013)\citenamefont
  {Stapfner}, \citenamefont {Ost}, \citenamefont {Hunger}, \citenamefont
  {Reichel}, \citenamefont {Favero},\ and\ \citenamefont
  {Weig}}]{Stapfner2013}%
  \BibitemOpen
  \bibfield  {author} {\bibinfo {author} {\bibfnamefont {S.}~\bibnamefont
  {Stapfner}}, \bibinfo {author} {\bibfnamefont {L.}~\bibnamefont {Ost}},
  \bibinfo {author} {\bibfnamefont {D.}~\bibnamefont {Hunger}}, \bibinfo
  {author} {\bibfnamefont {J.}~\bibnamefont {Reichel}}, \bibinfo {author}
  {\bibfnamefont {I.}~\bibnamefont {Favero}}, \ and\ \bibinfo {author}
  {\bibfnamefont {E.~M.}\ \bibnamefont {Weig}},\ }\href {\doibase
  10.1063/1.4802746} {\bibfield  {journal} {\bibinfo  {journal} {Appl. Phys.
  Lett.}\ }\textbf {\bibinfo {volume} {102}},\ \bibinfo {pages} {151910}
  (\bibinfo {year} {2013})}\BibitemShut {NoStop}%
\bibitem [{\citenamefont {Singh}\ \emph {et~al.}(2014)\citenamefont {Singh},
  \citenamefont {Bosman}, \citenamefont {Schneider}, \citenamefont {Blanter},
  \citenamefont {Castellanos-Gomez},\ and\ \citenamefont
  {{Steele}}}]{Singh2014}%
  \BibitemOpen
  \bibfield  {author} {\bibinfo {author} {\bibfnamefont {V.}~\bibnamefont
  {Singh}}, \bibinfo {author} {\bibfnamefont {S.~J.}\ \bibnamefont {Bosman}},
  \bibinfo {author} {\bibfnamefont {B.~H.}\ \bibnamefont {Schneider}}, \bibinfo
  {author} {\bibfnamefont {Y.~M.}\ \bibnamefont {Blanter}}, \bibinfo {author}
  {\bibfnamefont {A.}~\bibnamefont {Castellanos-Gomez}}, \ and\ \bibinfo
  {author} {\bibfnamefont {G.~A.}\ \bibnamefont {{Steele}}},\ }\href@noop {}
  {\bibfield  {journal} {\bibinfo  {journal} {Nat Nano}\ }\textbf {\bibinfo
  {volume} {9}},\ \bibinfo {pages} {820} (\bibinfo {year} {2014})}\BibitemShut
  {NoStop}%
\bibitem [{\citenamefont {Song}\ \emph {et~al.}(2014)\citenamefont {Song},
  \citenamefont {Oksanen}, \citenamefont {Li}, \citenamefont {Hakonen},\ and\
  \citenamefont {Sillanpa}}]{Song2014a}%
  \BibitemOpen
  \bibfield  {author} {\bibinfo {author} {\bibfnamefont {X.}~\bibnamefont
  {Song}}, \bibinfo {author} {\bibfnamefont {M.}~\bibnamefont {Oksanen}},
  \bibinfo {author} {\bibfnamefont {J.}~\bibnamefont {Li}}, \bibinfo {author}
  {\bibfnamefont {P.~J.}\ \bibnamefont {Hakonen}}, \ and\ \bibinfo {author}
  {\bibfnamefont {M.~A.}\ \bibnamefont {Sillanpa}},\ }\href
  {https://link.aps.org/doi/10.1103/PhysRevLett.113.027404} {\bibfield
  {journal} {\bibinfo  {journal} {Phys. Rev. Lett.}\ }\textbf {\bibinfo
  {volume} {113}},\ \bibinfo {pages} {027404} (\bibinfo {year}
  {2014})}\BibitemShut {NoStop}%
\bibitem [{\citenamefont {Weber}\ \emph {et~al.}(2014)\citenamefont {Weber},
  \citenamefont {G\"uttinger}, \citenamefont {Tsioutsios}, \citenamefont
  {Chang},\ and\ \citenamefont {Bachtold}}]{Weber2014}%
  \BibitemOpen
  \bibfield  {author} {\bibinfo {author} {\bibfnamefont {P.}~\bibnamefont
  {Weber}}, \bibinfo {author} {\bibfnamefont {J.}~\bibnamefont {G\"uttinger}},
  \bibinfo {author} {\bibfnamefont {I.}~\bibnamefont {Tsioutsios}}, \bibinfo
  {author} {\bibfnamefont {D.~E.}\ \bibnamefont {Chang}}, \ and\ \bibinfo
  {author} {\bibfnamefont {A.}~\bibnamefont {Bachtold}},\ }\href
  {http://dx.doi.org/10.1021/nl500879k} {\bibfield  {journal} {\bibinfo
  {journal} {Nano Lett.}\ }\textbf {\bibinfo {volume} {14}},\ \bibinfo {pages}
  {2854} (\bibinfo {year} {2014})}\BibitemShut {NoStop}%
\bibitem [{\citenamefont {Schneider}\ \emph {et~al.}(2014)\citenamefont
  {Schneider}, \citenamefont {Singh}, \citenamefont {Venstra}, \citenamefont
  {Meerwaldt},\ and\ \citenamefont {Steele}}]{Schneider2014}%
  \BibitemOpen
  \bibfield  {author} {\bibinfo {author} {\bibfnamefont {B.~H.}\ \bibnamefont
  {Schneider}}, \bibinfo {author} {\bibfnamefont {V.}~\bibnamefont {Singh}},
  \bibinfo {author} {\bibfnamefont {W.~J.}\ \bibnamefont {Venstra}}, \bibinfo
  {author} {\bibfnamefont {H.~B.}\ \bibnamefont {Meerwaldt}}, \ and\ \bibinfo
  {author} {\bibfnamefont {G.~A.}\ \bibnamefont {Steele}},\ }\href@noop {}
  {\bibfield  {journal} {\bibinfo  {journal} {Nat Commun}\ }\textbf {\bibinfo
  {volume} {5}},\ \bibinfo {pages} {5819} (\bibinfo {year} {2014})}\BibitemShut
  {NoStop}%
\bibitem [{\citenamefont {Nigues}\ \emph {et~al.}(2015)\citenamefont {Nigues},
  \citenamefont {Siria},\ and\ \citenamefont {Verlot}}]{Nigues2015}%
  \BibitemOpen
  \bibfield  {author} {\bibinfo {author} {\bibfnamefont {A.}~\bibnamefont
  {Nigues}}, \bibinfo {author} {\bibfnamefont {A.}~\bibnamefont {Siria}}, \
  and\ \bibinfo {author} {\bibfnamefont {P.}~\bibnamefont {Verlot}},\ }\href
  {http://dx.doi.org/10.1038/ncomms9104} {\bibfield  {journal} {\bibinfo
  {journal} {Nature Communications}\ }\textbf {\bibinfo {volume} {6}},\
  \bibinfo {pages} {8104} (\bibinfo {year} {2015})}\BibitemShut {NoStop}%
\bibitem [{\citenamefont {Cole}\ \emph {et~al.}(2015)\citenamefont {Cole},
  \citenamefont {Brawley}, \citenamefont {Adiga}, \citenamefont {De~Alba},
  \citenamefont {Parpia}, \citenamefont {Ilic}, \citenamefont {Craighead},\
  and\ \citenamefont {Bowen}}]{Cole2015}%
  \BibitemOpen
  \bibfield  {author} {\bibinfo {author} {\bibfnamefont {R.~M.}\ \bibnamefont
  {Cole}}, \bibinfo {author} {\bibfnamefont {G.~A.}\ \bibnamefont {Brawley}},
  \bibinfo {author} {\bibfnamefont {V.~P.}\ \bibnamefont {Adiga}}, \bibinfo
  {author} {\bibfnamefont {R.}~\bibnamefont {De~Alba}}, \bibinfo {author}
  {\bibfnamefont {J.~M.}\ \bibnamefont {Parpia}}, \bibinfo {author}
  {\bibfnamefont {B.}~\bibnamefont {Ilic}}, \bibinfo {author} {\bibfnamefont
  {H.~G.}\ \bibnamefont {Craighead}}, \ and\ \bibinfo {author} {\bibfnamefont
  {W.~P.}\ \bibnamefont {Bowen}},\ }\href
  {https://link.aps.org/doi/10.1103/PhysRevApplied.3.024004} {\bibfield
  {journal} {\bibinfo  {journal} {Phys. Rev. Applied}\ }\textbf {\bibinfo
  {volume} {3}},\ \bibinfo {pages} {024004} (\bibinfo {year}
  {2015})}\BibitemShut {NoStop}%
\bibitem [{\citenamefont {Tsioutsios}\ \emph {et~al.}(2017)\citenamefont
  {Tsioutsios}, \citenamefont {Tavernarakis}, \citenamefont {Osmond},
  \citenamefont {Verlot},\ and\ \citenamefont {Bachtold}}]{Tsioutsios2017}%
  \BibitemOpen
  \bibfield  {author} {\bibinfo {author} {\bibfnamefont {I.}~\bibnamefont
  {Tsioutsios}}, \bibinfo {author} {\bibfnamefont {A.}~\bibnamefont
  {Tavernarakis}}, \bibinfo {author} {\bibfnamefont {J.}~\bibnamefont
  {Osmond}}, \bibinfo {author} {\bibfnamefont {P.}~\bibnamefont {Verlot}}, \
  and\ \bibinfo {author} {\bibfnamefont {A.}~\bibnamefont {Bachtold}},\ }\href
  {\doibase 10.1021/acs.nanolett.6b05065} {\bibfield  {journal} {\bibinfo
  {journal} {Nano Lett.}\ }\textbf {\bibinfo {volume} {17}},\ \bibinfo {pages}
  {1748} (\bibinfo {year} {2017})}\BibitemShut {NoStop}%
\bibitem [{\citenamefont {{G\"uttinger}}\ \emph {et~al.}(2017)\citenamefont
  {{G\"uttinger}}, \citenamefont {{Noury}}, \citenamefont {{Weber}},
  \citenamefont {{Eriksson}}, \citenamefont {{Lagoin}}, \citenamefont
  {{Moser}}, \citenamefont {{Eichler}}, \citenamefont {{Wallraff}},
  \citenamefont {{Isacsson}},\ and\ \citenamefont
  {{Bachtold}}}]{Guettinger2017}%
  \BibitemOpen
  \bibfield  {author} {\bibinfo {author} {\bibfnamefont {J.}~\bibnamefont
  {{G\"uttinger}}}, \bibinfo {author} {\bibfnamefont {A.}~\bibnamefont
  {{Noury}}}, \bibinfo {author} {\bibfnamefont {P.}~\bibnamefont {{Weber}}},
  \bibinfo {author} {\bibfnamefont {A.~M.}\ \bibnamefont {{Eriksson}}},
  \bibinfo {author} {\bibfnamefont {C.}~\bibnamefont {{Lagoin}}}, \bibinfo
  {author} {\bibfnamefont {J.}~\bibnamefont {{Moser}}}, \bibinfo {author}
  {\bibfnamefont {C.}~\bibnamefont {{Eichler}}}, \bibinfo {author}
  {\bibfnamefont {A.}~\bibnamefont {{Wallraff}}}, \bibinfo {author}
  {\bibfnamefont {A.}~\bibnamefont {{Isacsson}}}, \ and\ \bibinfo {author}
  {\bibfnamefont {A.}~\bibnamefont {{Bachtold}}},\ }\href@noop {} {\bibfield
  {journal} {\bibinfo  {journal} {Nat. Nano.}\ }\textbf {\bibinfo {volume}
  {12}},\ \bibinfo {pages} {631} (\bibinfo {year} {2017})}\BibitemShut
  {NoStop}%
\bibitem [{\citenamefont {Tavernarakis}\ \emph {et~al.}(2018)\citenamefont
  {Tavernarakis}, \citenamefont {Stavrinadis}, \citenamefont {Nowak},
  \citenamefont {Tsioutsios}, \citenamefont {Bachtold},\ and\ \citenamefont
  {Verlot}}]{Tavernarakis2018}%
  \BibitemOpen
  \bibfield  {author} {\bibinfo {author} {\bibfnamefont {A.}~\bibnamefont
  {Tavernarakis}}, \bibinfo {author} {\bibfnamefont {A.}~\bibnamefont
  {Stavrinadis}}, \bibinfo {author} {\bibfnamefont {A.}~\bibnamefont {Nowak}},
  \bibinfo {author} {\bibfnamefont {I.}~\bibnamefont {Tsioutsios}}, \bibinfo
  {author} {\bibfnamefont {A.}~\bibnamefont {Bachtold}}, \ and\ \bibinfo
  {author} {\bibfnamefont {P.}~\bibnamefont {Verlot}},\ }\href
  {https://doi.org/10.1038/s41467-018-03097-z} {\bibfield  {journal} {\bibinfo
  {journal} {Nature Communications}\ }\textbf {\bibinfo {volume} {9}},\
  \bibinfo {pages} {662} (\bibinfo {year} {2018})}\BibitemShut {NoStop}%
\bibitem [{\citenamefont {Huang}\ \emph {et~al.}(2004)\citenamefont {Huang},
  \citenamefont {Woodson}, \citenamefont {Smalley},\ and\ \citenamefont
  {Liu}}]{Huang2004}%
  \BibitemOpen
  \bibfield  {author} {\bibinfo {author} {\bibfnamefont {S.}~\bibnamefont
  {Huang}}, \bibinfo {author} {\bibfnamefont {M.}~\bibnamefont {Woodson}},
  \bibinfo {author} {\bibfnamefont {R.}~\bibnamefont {Smalley}}, \ and\
  \bibinfo {author} {\bibfnamefont {J.}~\bibnamefont {Liu}},\ }\href {\doibase
  10.1021/nl049691d} {\bibfield  {journal} {\bibinfo  {journal} {Nano Lett.}\
  }\textbf {\bibinfo {volume} {4}},\ \bibinfo {pages} {1025} (\bibinfo {year}
  {2004})}\BibitemShut {NoStop}%
\bibitem [{\citenamefont {Jezouin}\ \emph {et~al.}(2013)\citenamefont
  {Jezouin}, \citenamefont {Parmentier}, \citenamefont {Anthore}, \citenamefont
  {Gennser}, \citenamefont {Cavanna}, \citenamefont {Jin},\ and\ \citenamefont
  {Pierre}}]{Jezouin2013}%
  \BibitemOpen
  \bibfield  {author} {\bibinfo {author} {\bibfnamefont {S.}~\bibnamefont
  {Jezouin}}, \bibinfo {author} {\bibfnamefont {F.~D.}\ \bibnamefont
  {Parmentier}}, \bibinfo {author} {\bibfnamefont {A.}~\bibnamefont {Anthore}},
  \bibinfo {author} {\bibfnamefont {U.}~\bibnamefont {Gennser}}, \bibinfo
  {author} {\bibfnamefont {A.}~\bibnamefont {Cavanna}}, \bibinfo {author}
  {\bibfnamefont {Y.}~\bibnamefont {Jin}}, \ and\ \bibinfo {author}
  {\bibfnamefont {F.}~\bibnamefont {Pierre}},\ }\href
  {http://science.sciencemag.org/content/342/6158/601.abstract} {\bibfield
  {journal} {\bibinfo  {journal} {Science}\ }\textbf {\bibinfo {volume}
  {342}},\ \bibinfo {pages} {601} (\bibinfo {year} {2013})}\BibitemShut
  {NoStop}%
\bibitem [{\citenamefont {Bocquillon}\ \emph {et~al.}(2013)\citenamefont
  {Bocquillon}, \citenamefont {Freulon}, \citenamefont {Berroir}, \citenamefont
  {Degiovanni}, \citenamefont {Placais}, \citenamefont {Cavanna}, \citenamefont
  {Jin},\ and\ \citenamefont {Feve}}]{Bocquillon2013}%
  \BibitemOpen
  \bibfield  {author} {\bibinfo {author} {\bibfnamefont {E.}~\bibnamefont
  {Bocquillon}}, \bibinfo {author} {\bibfnamefont {V.}~\bibnamefont {Freulon}},
  \bibinfo {author} {\bibfnamefont {J.-M.}\ \bibnamefont {Berroir}}, \bibinfo
  {author} {\bibfnamefont {P.}~\bibnamefont {Degiovanni}}, \bibinfo {author}
  {\bibfnamefont {B.}~\bibnamefont {Placais}}, \bibinfo {author} {\bibfnamefont
  {A.}~\bibnamefont {Cavanna}}, \bibinfo {author} {\bibfnamefont
  {Y.}~\bibnamefont {Jin}}, \ and\ \bibinfo {author} {\bibfnamefont
  {G.}~\bibnamefont {Feve}},\ }\href
  {http://science.sciencemag.org/content/early/2013/01/23/science.1232572.abstract}
  {\bibfield  {journal} {\bibinfo  {journal} {Science}\ } (\bibinfo {year}
  {2013})}\BibitemShut {NoStop}%
\bibitem [{\citenamefont {Jullien}\ \emph {et~al.}(2014)\citenamefont
  {Jullien}, \citenamefont {Roulleau}, \citenamefont {Roche}, \citenamefont
  {Cavanna}, \citenamefont {Jin},\ and\ \citenamefont {Glattli}}]{Jullien2014}%
  \BibitemOpen
  \bibfield  {author} {\bibinfo {author} {\bibfnamefont {T.}~\bibnamefont
  {Jullien}}, \bibinfo {author} {\bibfnamefont {P.}~\bibnamefont {Roulleau}},
  \bibinfo {author} {\bibfnamefont {B.}~\bibnamefont {Roche}}, \bibinfo
  {author} {\bibfnamefont {A.}~\bibnamefont {Cavanna}}, \bibinfo {author}
  {\bibfnamefont {Y.}~\bibnamefont {Jin}}, \ and\ \bibinfo {author}
  {\bibfnamefont {D.~C.}\ \bibnamefont {Glattli}},\ }\href
  {http://dx.doi.org/10.1038/nature13821} {\bibfield  {journal} {\bibinfo
  {journal} {Nature}\ }\textbf {\bibinfo {volume} {514}},\ \bibinfo {pages}
  {603} (\bibinfo {year} {2014})}\BibitemShut {NoStop}%
\bibitem [{\citenamefont {Dong}\ \emph {et~al.}(2014)\citenamefont {Dong},
  \citenamefont {Liang}, \citenamefont {Ferry}, \citenamefont {Cavanna},
  \citenamefont {Gennser}, \citenamefont {Couraud},\ and\ \citenamefont
  {Jin}}]{Dong2014}%
  \BibitemOpen
  \bibfield  {author} {\bibinfo {author} {\bibfnamefont {Q.}~\bibnamefont
  {Dong}}, \bibinfo {author} {\bibfnamefont {Y.~X.}\ \bibnamefont {Liang}},
  \bibinfo {author} {\bibfnamefont {D.}~\bibnamefont {Ferry}}, \bibinfo
  {author} {\bibfnamefont {A.}~\bibnamefont {Cavanna}}, \bibinfo {author}
  {\bibfnamefont {U.}~\bibnamefont {Gennser}}, \bibinfo {author} {\bibfnamefont
  {L.}~\bibnamefont {Couraud}}, \ and\ \bibinfo {author} {\bibfnamefont
  {Y.}~\bibnamefont {Jin}},\ }\href {\doibase 10.1063/1.4887368} {\bibfield
  {journal} {\bibinfo  {journal} {Appl. Phys. Lett.}\ }\textbf {\bibinfo
  {volume} {105}},\ \bibinfo {pages} {013504} (\bibinfo {year}
  {2014})}\BibitemShut {NoStop}%
\bibitem [{\citenamefont {Song}\ \emph {et~al.}(2012)\citenamefont {Song},
  \citenamefont {Oksanen}, \citenamefont {Sillanpa}, \citenamefont {Craighead},
  \citenamefont {Parpia},\ and\ \citenamefont {Hakonen}}]{Song2012}%
  \BibitemOpen
  \bibfield  {author} {\bibinfo {author} {\bibfnamefont {X.}~\bibnamefont
  {Song}}, \bibinfo {author} {\bibfnamefont {M.}~\bibnamefont {Oksanen}},
  \bibinfo {author} {\bibfnamefont {M.~A.}\ \bibnamefont {Sillanpa}}, \bibinfo
  {author} {\bibfnamefont {H.~G.}\ \bibnamefont {Craighead}}, \bibinfo {author}
  {\bibfnamefont {J.~M.}\ \bibnamefont {Parpia}}, \ and\ \bibinfo {author}
  {\bibfnamefont {P.~J.}\ \bibnamefont {Hakonen}},\ }\href {\doibase
  10.1021/nl203305q} {\bibfield  {journal} {\bibinfo  {journal} {Nano Lett.}\
  }\textbf {\bibinfo {volume} {12}},\ \bibinfo {pages} {198} (\bibinfo {year}
  {2012})}\BibitemShut {NoStop}%
\bibitem [{\citenamefont {Weber}\ \emph {et~al.}(2016)\citenamefont {Weber},
  \citenamefont {Guttinger}, \citenamefont {Noury}, \citenamefont
  {Vergara-Cruz},\ and\ \citenamefont {Bachtold}}]{Weber2016}%
  \BibitemOpen
  \bibfield  {author} {\bibinfo {author} {\bibfnamefont {P.}~\bibnamefont
  {Weber}}, \bibinfo {author} {\bibfnamefont {J.}~\bibnamefont {Guttinger}},
  \bibinfo {author} {\bibfnamefont {A.}~\bibnamefont {Noury}}, \bibinfo
  {author} {\bibfnamefont {J.}~\bibnamefont {Vergara-Cruz}}, \ and\ \bibinfo
  {author} {\bibfnamefont {A.}~\bibnamefont {Bachtold}},\ }\href {\doibase
  10.1038/ncomms12496} {\bibfield  {journal} {\bibinfo  {journal} {Nature
  Communications}\ }\textbf {\bibinfo {volume} {7}},\ \bibinfo {pages} {12496}
  (\bibinfo {year} {2016})}\BibitemShut {NoStop}%
\bibitem [{\citenamefont {Teufel}\ \emph {et~al.}(2009)\citenamefont {Teufel},
  \citenamefont {Donner}, \citenamefont {Castellanos-Beltran}, \citenamefont
  {Harlow},\ and\ \citenamefont {Lehnert}}]{Teufel2009}%
  \BibitemOpen
  \bibfield  {author} {\bibinfo {author} {\bibfnamefont {J.~D.}\ \bibnamefont
  {Teufel}}, \bibinfo {author} {\bibfnamefont {T.}~\bibnamefont {Donner}},
  \bibinfo {author} {\bibfnamefont {M.~A.}\ \bibnamefont
  {Castellanos-Beltran}}, \bibinfo {author} {\bibfnamefont {J.~W.}\
  \bibnamefont {Harlow}}, \ and\ \bibinfo {author} {\bibfnamefont {K.~W.}\
  \bibnamefont {Lehnert}},\ }\href@noop {} {\bibfield  {journal} {\bibinfo
  {journal} {Nature Nanotech.}\ }\textbf {\bibinfo {volume} {4}},\ \bibinfo
  {pages} {820} (\bibinfo {year} {2009})}\BibitemShut {NoStop}%
\bibitem [{\citenamefont {Heritier}\ \emph {et~al.}(2018)\citenamefont
  {Heritier}, \citenamefont {Eichler}, \citenamefont {Pan}, \citenamefont
  {Grob}, \citenamefont {Shorubalko}, \citenamefont {Krass}, \citenamefont
  {Tao},\ and\ \citenamefont {Degen}}]{Heritier2018}%
  \BibitemOpen
  \bibfield  {author} {\bibinfo {author} {\bibfnamefont {M.}~\bibnamefont
  {Heritier}}, \bibinfo {author} {\bibfnamefont {A.}~\bibnamefont {Eichler}},
  \bibinfo {author} {\bibfnamefont {Y.}~\bibnamefont {Pan}}, \bibinfo {author}
  {\bibfnamefont {U.}~\bibnamefont {Grob}}, \bibinfo {author} {\bibfnamefont
  {I.}~\bibnamefont {Shorubalko}}, \bibinfo {author} {\bibfnamefont {M.~D.}\
  \bibnamefont {Krass}}, \bibinfo {author} {\bibfnamefont {Y.}~\bibnamefont
  {Tao}}, \ and\ \bibinfo {author} {\bibfnamefont {C.~L.}\ \bibnamefont
  {Degen}},\ }\href {\doibase 10.1021/acs.nanolett.7b05035} {\bibfield
  {journal} {\bibinfo  {journal} {Nano Lett.}\ }\textbf {\bibinfo {volume}
  {18}},\ \bibinfo {pages} {1814} (\bibinfo {year} {2018})}\BibitemShut
  {NoStop}%
\bibitem [{\citenamefont {Reinhardt}\ \emph {et~al.}(2016)\citenamefont
  {Reinhardt}, \citenamefont {Muller}, \citenamefont {Bourassa},\ and\
  \citenamefont {Sankey}}]{Reinhardt2016}%
  \BibitemOpen
  \bibfield  {author} {\bibinfo {author} {\bibfnamefont {C.}~\bibnamefont
  {Reinhardt}}, \bibinfo {author} {\bibfnamefont {T.}~\bibnamefont {Muller}},
  \bibinfo {author} {\bibfnamefont {A.}~\bibnamefont {Bourassa}}, \ and\
  \bibinfo {author} {\bibfnamefont {J.~C.}\ \bibnamefont {Sankey}},\ }\href
  {https://link.aps.org/doi/10.1103/PhysRevX.6.021001} {\bibfield  {journal}
  {\bibinfo  {journal} {Phys. Rev. X}\ }\textbf {\bibinfo {volume} {6}},\
  \bibinfo {pages} {021001} (\bibinfo {year} {2016})}\BibitemShut {NoStop}%
\bibitem [{\citenamefont {Nichol}\ \emph {et~al.}(2012)\citenamefont {Nichol},
  \citenamefont {Hemesath}, \citenamefont {Lauhon},\ and\ \citenamefont
  {Budakian}}]{Nichol2012}%
  \BibitemOpen
  \bibfield  {author} {\bibinfo {author} {\bibfnamefont {J.~M.}\ \bibnamefont
  {Nichol}}, \bibinfo {author} {\bibfnamefont {E.~R.}\ \bibnamefont
  {Hemesath}}, \bibinfo {author} {\bibfnamefont {L.~J.}\ \bibnamefont
  {Lauhon}}, \ and\ \bibinfo {author} {\bibfnamefont {R.}~\bibnamefont
  {Budakian}},\ }\href {https://link.aps.org/doi/10.1103/PhysRevB.85.054414}
  {\bibfield  {journal} {\bibinfo  {journal} {Phys. Rev. B}\ }\textbf {\bibinfo
  {volume} {85}},\ \bibinfo {pages} {054414} (\bibinfo {year}
  {2012})}\BibitemShut {NoStop}%
\bibitem [{\citenamefont {Wang}\ and\ \citenamefont
  {Pistolesi}(2017)}]{Wang2017a}%
  \BibitemOpen
  \bibfield  {author} {\bibinfo {author} {\bibfnamefont {Y.}~\bibnamefont
  {Wang}}\ and\ \bibinfo {author} {\bibfnamefont {F.}~\bibnamefont
  {Pistolesi}},\ }\href {https://link.aps.org/doi/10.1103/PhysRevB.95.035410}
  {\bibfield  {journal} {\bibinfo  {journal} {Phys. Rev. B}\ }\textbf {\bibinfo
  {volume} {95}},\ \bibinfo {pages} {035410} (\bibinfo {year}
  {2017})}\BibitemShut {NoStop}%
\bibitem [{\citenamefont {Degen}\ \emph {et~al.}(2009)\citenamefont {Degen},
  \citenamefont {Poggio}, \citenamefont {Mamin}, \citenamefont {Rettner},\ and\
  \citenamefont {Rugar}}]{Degen2009}%
  \BibitemOpen
  \bibfield  {author} {\bibinfo {author} {\bibfnamefont {C.~L.}\ \bibnamefont
  {Degen}}, \bibinfo {author} {\bibfnamefont {M.}~\bibnamefont {Poggio}},
  \bibinfo {author} {\bibfnamefont {H.~J.}\ \bibnamefont {Mamin}}, \bibinfo
  {author} {\bibfnamefont {C.~T.}\ \bibnamefont {Rettner}}, \ and\ \bibinfo
  {author} {\bibfnamefont {D.}~\bibnamefont {Rugar}},\ }\href
  {http://www.pnas.org/content/106/5/1313.abstract} {\bibfield  {journal}
  {\bibinfo  {journal} {Proc Natl Acad Sci USA}\ }\textbf {\bibinfo {volume}
  {106}},\ \bibinfo {pages} {1313} (\bibinfo {year} {2009})}\BibitemShut
  {NoStop}%
\bibitem [{\citenamefont {Mamin}\ \emph {et~al.}(2007)\citenamefont {Mamin},
  \citenamefont {Poggio}, \citenamefont {Degen},\ and\ \citenamefont
  {Rugar}}]{Mamin2007a}%
  \BibitemOpen
  \bibfield  {author} {\bibinfo {author} {\bibfnamefont {H.~J.}\ \bibnamefont
  {Mamin}}, \bibinfo {author} {\bibfnamefont {M.}~\bibnamefont {Poggio}},
  \bibinfo {author} {\bibfnamefont {C.~L.}\ \bibnamefont {Degen}}, \ and\
  \bibinfo {author} {\bibfnamefont {D.}~\bibnamefont {Rugar}},\ }\href
  {http://dx.doi.org/10.1038/nnano.2007.105} {\bibfield  {journal} {\bibinfo
  {journal} {Nature Nanotechnology}\ }\textbf {\bibinfo {volume} {2}},\
  \bibinfo {pages} {301} (\bibinfo {year} {2007})}\BibitemShut {NoStop}%
\bibitem [{\citenamefont {Nichol}\ \emph {et~al.}(2013)\citenamefont {Nichol},
  \citenamefont {Naibert}, \citenamefont {Hemesath}, \citenamefont {Lauhon},\
  and\ \citenamefont {Budakian}}]{Nichol2013}%
  \BibitemOpen
  \bibfield  {author} {\bibinfo {author} {\bibfnamefont {J.~M.}\ \bibnamefont
  {Nichol}}, \bibinfo {author} {\bibfnamefont {T.~R.}\ \bibnamefont {Naibert}},
  \bibinfo {author} {\bibfnamefont {E.~R.}\ \bibnamefont {Hemesath}}, \bibinfo
  {author} {\bibfnamefont {L.~J.}\ \bibnamefont {Lauhon}}, \ and\ \bibinfo
  {author} {\bibfnamefont {R.}~\bibnamefont {Budakian}},\ }\href
  {https://link.aps.org/doi/10.1103/PhysRevX.3.031016} {\bibfield  {journal}
  {\bibinfo  {journal} {Phys. Rev. X}\ }\textbf {\bibinfo {volume} {3}},\
  \bibinfo {pages} {031016} (\bibinfo {year} {2013})}\BibitemShut {NoStop}%
\bibitem [{\citenamefont {Armour}\ \emph {et~al.}(2004)\citenamefont {Armour},
  \citenamefont {Blencowe},\ and\ \citenamefont {Zhang}}]{Armour2004}%
  \BibitemOpen
  \bibfield  {author} {\bibinfo {author} {\bibfnamefont {A.~D.}\ \bibnamefont
  {Armour}}, \bibinfo {author} {\bibfnamefont {M.~P.}\ \bibnamefont
  {Blencowe}}, \ and\ \bibinfo {author} {\bibfnamefont {Y.}~\bibnamefont
  {Zhang}},\ }\href@noop {} {\bibfield  {journal} {\bibinfo  {journal} {Phys.
  Rev. B}\ }\textbf {\bibinfo {volume} {69}},\ \bibinfo {pages} {125313}
  (\bibinfo {year} {2004})}\BibitemShut {NoStop}%
\bibitem [{\citenamefont {Clerk}\ and\ \citenamefont
  {Bennett}(2005)}]{Clerk2005}%
  \BibitemOpen
  \bibfield  {author} {\bibinfo {author} {\bibfnamefont {A.~A.}\ \bibnamefont
  {Clerk}}\ and\ \bibinfo {author} {\bibfnamefont {S.}~\bibnamefont
  {Bennett}},\ }\href@noop {} {\bibfield  {journal} {\bibinfo  {journal} {New
  J. Phys.}\ }\textbf {\bibinfo {volume} {7}},\ \bibinfo {pages} {238}
  (\bibinfo {year} {2005})}\BibitemShut {NoStop}%
\bibitem [{\citenamefont {Naik}\ \emph {et~al.}(2006)\citenamefont {Naik},
  \citenamefont {Buu}, \citenamefont {LaHaye}, \citenamefont {Armour},
  \citenamefont {Clerk}, \citenamefont {Blencowe},\ and\ \citenamefont
  {Schwab}}]{Naik2006}%
  \BibitemOpen
  \bibfield  {author} {\bibinfo {author} {\bibfnamefont {A.}~\bibnamefont
  {Naik}}, \bibinfo {author} {\bibfnamefont {O.}~\bibnamefont {Buu}}, \bibinfo
  {author} {\bibfnamefont {M.~D.}\ \bibnamefont {LaHaye}}, \bibinfo {author}
  {\bibfnamefont {A.~D.}\ \bibnamefont {Armour}}, \bibinfo {author}
  {\bibfnamefont {A.~A.}\ \bibnamefont {Clerk}}, \bibinfo {author}
  {\bibfnamefont {M.~P.}\ \bibnamefont {Blencowe}}, \ and\ \bibinfo {author}
  {\bibfnamefont {K.~C.}\ \bibnamefont {Schwab}},\ }\href@noop {} {\bibfield
  {journal} {\bibinfo  {journal} {Nature}\ }\textbf {\bibinfo {volume} {443}},\
  \bibinfo {pages} {193} (\bibinfo {year} {2006})}\BibitemShut {NoStop}%
\bibitem [{\citenamefont {Micchi}\ \emph {et~al.}(2015)\citenamefont {Micchi},
  \citenamefont {Avriller},\ and\ \citenamefont {Pistolesi}}]{Micchi2015}%
  \BibitemOpen
  \bibfield  {author} {\bibinfo {author} {\bibfnamefont {G.}~\bibnamefont
  {Micchi}}, \bibinfo {author} {\bibfnamefont {R.}~\bibnamefont {Avriller}}, \
  and\ \bibinfo {author} {\bibfnamefont {F.}~\bibnamefont {Pistolesi}},\
  }\href@noop {} {\bibfield  {journal} {\bibinfo  {journal} {Phys. Rev. Lett.}\
  }\textbf {\bibinfo {volume} {115}},\ \bibinfo {pages} {206802} (\bibinfo
  {year} {2015})}\BibitemShut {NoStop}%
\bibitem [{\citenamefont {Pistolesi}\ and\ \citenamefont
  {Labarthe}(2007)}]{Pistolesi2007}%
  \BibitemOpen
  \bibfield  {author} {\bibinfo {author} {\bibfnamefont {F.}~\bibnamefont
  {Pistolesi}}\ and\ \bibinfo {author} {\bibfnamefont {S.}~\bibnamefont
  {Labarthe}},\ }\href {\doibase 10.1103/PhysRevB.76.165317} {\bibfield
  {journal} {\bibinfo  {journal} {Physical Review B}\ }\textbf {\bibinfo
  {volume} {76}},\ \bibinfo {pages} {165317} (\bibinfo {year}
  {2007})}\BibitemShut {NoStop}%
\bibitem [{\citenamefont {Weick}\ \emph {et~al.}(2011)\citenamefont {Weick},
  \citenamefont {von Oppen},\ and\ \citenamefont {Pistolesi}}]{Weick2011}%
  \BibitemOpen
  \bibfield  {author} {\bibinfo {author} {\bibfnamefont {G.}~\bibnamefont
  {Weick}}, \bibinfo {author} {\bibfnamefont {F.}~\bibnamefont {von Oppen}}, \
  and\ \bibinfo {author} {\bibfnamefont {F.}~\bibnamefont {Pistolesi}},\ }\href
  {\doibase 10.1103/PhysRevB.83.035420} {\bibfield  {journal} {\bibinfo
  {journal} {Phys. Rev. B}\ }\textbf {\bibinfo {volume} {83}},\ \bibinfo
  {pages} {035420} (\bibinfo {year} {2011})}\BibitemShut {NoStop}%
\bibitem [{\citenamefont {Micchi}\ \emph {et~al.}(2016)\citenamefont {Micchi},
  \citenamefont {Avriller},\ and\ \citenamefont {Pistolesi}}]{Micchi2016}%
  \BibitemOpen
  \bibfield  {author} {\bibinfo {author} {\bibfnamefont {G.}~\bibnamefont
  {Micchi}}, \bibinfo {author} {\bibfnamefont {R.}~\bibnamefont {Avriller}}, \
  and\ \bibinfo {author} {\bibfnamefont {F.}~\bibnamefont {Pistolesi}},\ }\href
  {https://link.aps.org/doi/10.1103/PhysRevB.94.125417} {\bibfield  {journal}
  {\bibinfo  {journal} {Phys. Rev. B}\ }\textbf {\bibinfo {volume} {94}},\
  \bibinfo {pages} {125417} (\bibinfo {year} {2016})}\BibitemShut {NoStop}%
\bibitem [{\citenamefont {Zippilli}\ \emph {et~al.}(2009)\citenamefont
  {Zippilli}, \citenamefont {Morigi},\ and\ \citenamefont
  {Bachtold}}]{Zippilli2009}%
  \BibitemOpen
  \bibfield  {author} {\bibinfo {author} {\bibfnamefont {S.}~\bibnamefont
  {Zippilli}}, \bibinfo {author} {\bibfnamefont {G.}~\bibnamefont {Morigi}}, \
  and\ \bibinfo {author} {\bibfnamefont {A.}~\bibnamefont {Bachtold}},\ }\href
  {https://link.aps.org/doi/10.1103/PhysRevLett.102.096804} {\bibfield
  {journal} {\bibinfo  {journal} {Phys. Rev. Lett.}\ }\textbf {\bibinfo
  {volume} {102}},\ \bibinfo {pages} {096804} (\bibinfo {year}
  {2009})}\BibitemShut {NoStop}%
\bibitem [{\citenamefont {Santandrea}\ \emph {et~al.}(2011)\citenamefont
  {Santandrea}, \citenamefont {Gorelik}, \citenamefont {Shekhter},\ and\
  \citenamefont {Jonson}}]{Santandrea2011}%
  \BibitemOpen
  \bibfield  {author} {\bibinfo {author} {\bibfnamefont {F.}~\bibnamefont
  {Santandrea}}, \bibinfo {author} {\bibfnamefont {L.~Y.}\ \bibnamefont
  {Gorelik}}, \bibinfo {author} {\bibfnamefont {R.~I.}\ \bibnamefont
  {Shekhter}}, \ and\ \bibinfo {author} {\bibfnamefont {M.}~\bibnamefont
  {Jonson}},\ }\href@noop {} {\bibfield  {journal} {\bibinfo  {journal} {Phys.
  Rev. Lett.}\ }\textbf {\bibinfo {volume} {106}},\ \bibinfo {pages} {186803}
  (\bibinfo {year} {2011})}\BibitemShut {NoStop}%
\bibitem [{\citenamefont {Stadler}\ \emph {et~al.}(2014)\citenamefont
  {Stadler}, \citenamefont {Belzig},\ and\ \citenamefont
  {Rastelli}}]{Stadler2014}%
  \BibitemOpen
  \bibfield  {author} {\bibinfo {author} {\bibfnamefont {P.}~\bibnamefont
  {Stadler}}, \bibinfo {author} {\bibfnamefont {W.}~\bibnamefont {Belzig}}, \
  and\ \bibinfo {author} {\bibfnamefont {G.}~\bibnamefont {Rastelli}},\ }\href
  {https://link.aps.org/doi/10.1103/PhysRevLett.113.047201} {\bibfield
  {journal} {\bibinfo  {journal} {Phys. Rev. Lett.}\ }\textbf {\bibinfo
  {volume} {113}},\ \bibinfo {pages} {047201} (\bibinfo {year}
  {2014})}\BibitemShut {NoStop}%
\bibitem [{\citenamefont {Arrachea}\ \emph {et~al.}(2014)\citenamefont
  {Arrachea}, \citenamefont {Bode},\ and\ \citenamefont {von
  Oppen}}]{Arrachea2014}%
  \BibitemOpen
  \bibfield  {author} {\bibinfo {author} {\bibfnamefont {L.}~\bibnamefont
  {Arrachea}}, \bibinfo {author} {\bibfnamefont {N.}~\bibnamefont {Bode}}, \
  and\ \bibinfo {author} {\bibfnamefont {F.}~\bibnamefont {von Oppen}},\ }\href
  {https://link.aps.org/doi/10.1103/PhysRevB.90.125450} {\bibfield  {journal}
  {\bibinfo  {journal} {Phys. Rev. B}\ }\textbf {\bibinfo {volume} {90}},\
  \bibinfo {pages} {125450} (\bibinfo {year} {2014})}\BibitemShut {NoStop}%
\bibitem [{\citenamefont {Stadler}\ \emph {et~al.}(2016)\citenamefont
  {Stadler}, \citenamefont {Belzig},\ and\ \citenamefont
  {Rastelli}}]{Stadler2016}%
  \BibitemOpen
  \bibfield  {author} {\bibinfo {author} {\bibfnamefont {P.}~\bibnamefont
  {Stadler}}, \bibinfo {author} {\bibfnamefont {W.}~\bibnamefont {Belzig}}, \
  and\ \bibinfo {author} {\bibfnamefont {G.}~\bibnamefont {Rastelli}},\ }\href
  {https://link.aps.org/doi/10.1103/PhysRevLett.117.197202} {\bibfield
  {journal} {\bibinfo  {journal} {Phys. Rev. Lett.}\ }\textbf {\bibinfo
  {volume} {117}},\ \bibinfo {pages} {197202} (\bibinfo {year}
  {2016})}\BibitemShut {NoStop}%
\end{thebibliography}

\end{document}